\documentclass[pdflatex,sn-mathphys-num]{sn-jnl}

\usepackage{graphicx}%
\usepackage{multirow}%
\usepackage{amsmath,amssymb,amsfonts}%
\usepackage{amsthm}%
\usepackage{mathrsfs}%
\usepackage[title]{appendix}%
\usepackage{xcolor}%
\usepackage{textcomp}%
\usepackage{manyfoot}%
\usepackage{booktabs}%
\usepackage{algorithm}%
\usepackage{algorithmicx}%
\usepackage{algpseudocode}%
\usepackage{listings}%

\usepackage{xr}

\theoremstyle{thmstyleone}%
\theoremstyle{thmstyletwo}%

\theoremstyle{thmstylethree}%

\begin{document}
\title{Ionic glass formers show an inverted relation between fragility and relaxation broadness}


\author*[1]{\fnm{Sophie G.M.} \spfx{van} \sur{Lange}}\email{sophie.vanlange@wur.nl}
\author[1]{\fnm{Diane W.} \spfx{te} \sur{Brake}}
\author[1]{\fnm{Eline F.} \sur{Brink}}
\author[1]{\fnm{Jochem} \sur{Pees}}
\author[1]{\fnm{Mathilde M.} \spfx{van} \sur{Nieuwenhuijzen}}
\author[2]{\fnm{Nayan} \sur{Vengallur}}
\author[3]{\fnm{Alessio} \sur{Zaccone}}
\author[2]{\fnm{Andrea} \sur{Giuntoli}}
\author[4]{\fnm{Joris} \sur{Sprakel}}
\author*[1]{\fnm{Jasper} \spfx{van der} \sur{Gucht}}\email{jasper.vandergucht@wur.nl}



\affil*[1]{\orgdiv{Physical Chemistry and Soft Matter}, \orgname{Wageningen University and Research}, \orgaddress{\street{Stippeneng 4}, \city{Wageningen}, \postcode{6708 WE}, \country{The Netherlands}}}
\affil[2]{\orgdiv{Zernike Institute for Advanced Materials}, \orgname{University of Groningen}, \orgaddress{\street{Nijenborgh 3}, \city{Groningen}, \postcode{9747 AG}, \country{The Netherlands}}}
\affil[3]{\orgdiv{Department of Physics “A. Pontremoli”}, \orgname{University of Milan}, \orgaddress{\street{Via Celoria 16}, \city{Milan}, \postcode{20133}, \country{Italy}}}
\affil[4]{\orgdiv{Laboratory of Biochemistry}, \orgname{Wageningen University and Research}, \orgaddress{\street{Stippeneng 4}, \city{Wageningen}, \postcode{6708WE}, \country{The Netherlands}}}


\abstract{Supercooled liquids undergo a rapid change in dynamics as they are cooled to their glass transition temperature and turn from a flowing liquid into an amorphous solid. Depending on how steeply the viscosity changes with temperature around the glass transition, glass formers are classified as strong or fragile\cite{angell1995formation, angell1991relaxation}. An empirical relation exists between the fragility of the liquid and the broadness of its relaxation spectrum\cite{boehmer1993nonexponential, plazek1991correlation}. However, the microscopic origins of this correlation remain unclear and its generality has been debated\cite{hong2011there, heuer2008exploring, dyre2007ten}. Here, we demonstrate that this relationship is inverted in organic materials with ionic interactions. We introduce a novel class of materials consisting of highly charged hydrophobic polymers cross-linked via moderated ionic interactions, and show that these combine a strong glass transition with an unusually broad mechanical relaxation spectrum. By surveying a large variety of ionic liquids, polymerized ionic liquids, and ionomers, we show that all these charged materials follow a trend between fragility and relaxation broadness that is opposite to that of non-charged materials. This finding suggests a special role of long-ranged ionic interactions in vitrification and opens up a route toward developing new materials that combine the processability of strong glass formers with the mechanical dissipation of polymers.}

\keywords{Fragility, Ionic glass formers, Non-exponential relaxation, polymer glasses}



\maketitle


The thermomechanical properties of silica glass have historically enabled glass-makers to shape, manipulate, and blow molten glass on metal rods without it flowing off. Inorganic network glasses such as SiO\textsubscript{2} exhibit these unique and desirable processing properties because of their `strong' glass-forming nature: their viscosity decreases gradually over a wide temperature range upon heating above the glass transition temperature ($T_g$). By contrast, nearly all organic glass formers and polymers are `fragile' glass formers, showing a very abrupt decrease in viscosity within a narrow temperature window above the glass transition. This makes it difficult to process these materials without a mold or precise temperature control. The classification scheme for glass formers as either strong or fragile is based on the extent to which their structural relaxation time $\tau$ deviates from Arrhenius behavior. This can be effectively captured by a parameter called the fragility index\cite{angell1995formation, angell1991relaxation},

\begin{equation}
    m = \left. \frac{d \log_{10} \tau}{d \left(T_g/T \right)} \right|_{T = T_g}.
    \label{ch4eq1}
\end{equation}

\noindent Strong glass formers, such as silica, other inorganic network glasses, and more recently discovered organic vitrimer networks closely follow the Arrhenius relation \cite{montarnal2011silica, denissen2016vitrimers, ciarella2019understanding}, which corresponds to a low fragility index, $m\sim 16-25$. 
Most molecular liquids, however, deviate strongly from the Arrhenius relation and display a much steeper temperature dependence near $T_g$, leading to a higher fragility index $m\sim 50-80$, and for most polymers $m>100$\cite{dalle2016many}. 

While the concept of fragility plays a crucial role in the phenomenology of the glass transition, its microscopic origin remains poorly understood. Nevertheless, a large number of experimental studies have provided valuable insights in the relation between fragility and the microstructure of the material. It is well established that strong glass formation often requires a network structure, where relaxation is governed by the thermally activated dissociation of directional (covalent) bonds \cite{saika2001fragile, debenedetti2001supercooled, sastry1998signatures,hall2003microscopic, ciarella2019understanding}. Fragile glass formers, on the other hand, are characterized by non-directional interactions, such as van der Waals or excluded volume interactions, and dynamics that are governed by cooperative molecular rearrangements \cite{debenedetti2001supercooled, ruocco2004landscapes, pazmino2018string, furukawa2016significant, betancourt2013fragility, hall2003microscopic, vilgis1993strong, krausser2015interatomic}. 
This cooperativity in the dynamics is typically associated with non-exponential relaxation processes, commonly described by the Kohlrausch–Williams–Watts relation\cite{kohlrausch1854theorie, williams1970non},

\begin{equation}
    \phi(t) \sim \exp\left[-\left(t/\tau\right)^\beta\right]
    \label{ch4eq2}
\end{equation}

\noindent where $\phi(t)$ is the response function measured with, e.g., rheology, dielectric relaxation, or dynamic scattering techniques. The stretch exponent $\beta$ quantifies the deviation from single-exponential decay and is related to the width of the relaxation spectrum. For $\beta \approx1$, the relaxation is governed by a single molecular process, while for $\beta<1$  the relaxation spectrum is broader and  governed by cooperative effects or dynamic heterogeneities. An extensive survey of a large number of different glass formers has indicated a clear correlation between the non-exponentiality of the relaxation of a glass former and its fragility\cite{boehmer1993nonexponential, plazek1991correlation}. Strong glass formers, such as silica, and other network glasses, have an almost single-exponential relaxation function, with a characteristic relaxation time that is determined by the lifetime of the covalent bonds. More fragile glasses have a broader relaxation spectrum, corresponding to a smaller value of the stretch exponent. Various theoretical models have been proposed to interpret this correlation in terms of the shape of the potential energy landscape\cite{zhang2022angell, gupta2008two}, spatial heterogeneities\cite{xia2001microscopic}, or fluctuations in the local coordination number\cite{ikeda2010correlation, vilgis1993strong}, but their validity remains debated\cite{hong2011there, heuer2008exploring, dyre2007ten}.

The empirical correlation between fragility and non-exponential relaxation constrains the development of new materials that integrate the processability of strong glass formers with the mechanical dissipation and energy absorption properties associated with a broad relaxation spectrum. Here we present a novel class of charged polymer materials that overcomes this limitation, displaying both a strong glass transition and an exceptionally broad relaxation spectrum. By analyzing literature data for other classes of  charged glass formers, we show that this unusual behaviour is related to the ionic interactions in these materials, which give rise to a trend between fragility and non-exponentiality that is opposite to that for non-charged materials.



\subsection*{Compleximers: ionically cross-linked polymers with a broad thermal transition}

\begin{figure}[t!]
\begin{center}
\includegraphics[width=\textwidth]{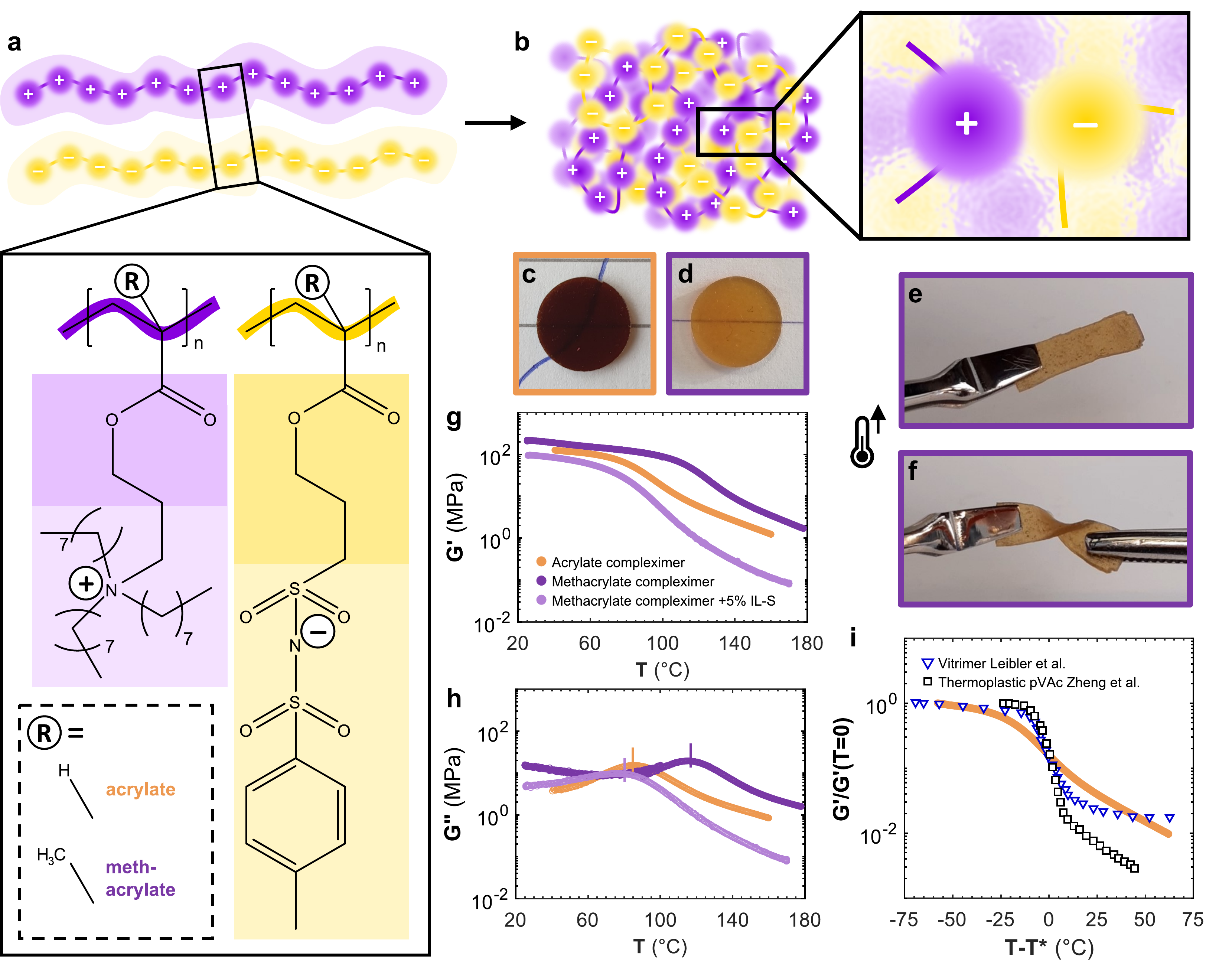}
\caption{\textbf{Hydrophobic polyelectrolytes form thermally deformable compleximers.} \textbf{(a)} The chemical structure of hydrophobic acrylate- and methacrylate-based polyelectrolytes, with charges that are screened by bulky attenuator chains. \textbf{(b)} Schematic of complex. Acrylate \textbf{(c)}, and methacrylate \textbf{(d)} compleximer powder can be reversibly hot-pressed into transparent samples. Methacrylate compleximer \textbf{(e)} can be easily reshaped upon heating with a heat gun \textbf{(f)}. \textbf{(g)} The dynamic mechanical profiles of compleximers show a gradual decrease in the  storage modulus ($G’$) with temperature. \textbf{(h)} The broadness of this transition is also observed in the loss modulus ($G”$), which marks a glass transition around 84 \textdegree C and 116 \textdegree C for acrylate and methacrylate compleximers, respectively. Plasticization with 5\% ionic liquid reduces the $T_g$ of the methacrylate compleximer to 80 \textdegree C. \textbf{(i)} The thermal transition of acrylate-based compleximer (orange) compared to two other thermoplastics, polyvinyl acetate (pVAc) and a vitrimeric hydroxy-ester network, which both show much steeper drops in the modulus with temperature. For clarity, we normalized the plateau modulus with its low-temperature value and shifted the curves horizontally with respect to the steepest point in the thermal transition (T*).}
\label{ch4fig1}
\end{center}
\end{figure}


In both silica \cite{ojavan2021on} and vitrimers \cite{yang2019detecting}, the strong nature of the glass transition arises from their network structure, where bonds become dynamic above $T_g$. Although this behavior has been primarily associated with specific covalent bonds, we hypothesized that similar properties can be achieved through ionic interactions. Indeed, higher charge densities have previously been linked to reduced fragility in ionic liquids and polymerized ionic liquids.\cite{nakamura2013viscoelastic, ueno2012protic}. 
Introducing ionic interactions into organic polymeric materials can be accomplished by forming complexes between oppositely charged polymers\cite{michaels1965polyelectrolyte}. Such complexes are usually extremely sensitive to water: they become very soft when hydrated \cite{wang2014the}, but extremely brittle in dry conditions, lacking an accessible glass transition and making them difficult to process \cite{zhang2016effect}. Recently, we demonstrated that these problems can be solved by attaching extended hydrophobic tails to the charged monomers\cite{vanlange2024moderated}, which reduces the proximity of the ionic groups. This molecular structure, inspired by ionic liquids \cite{gebbie2017long}, moderates the strength of the ionic interactions, making the materials malleable at elevated temperatures, while also making the polymers impervious to moisture. The resulting materials, which we called ‘compleximers’, exhibited properties similar to those of vitrimers; they combine the solvent resistance of thermoset polymers with the processability and recyclability of silica glass. 

Our first version of compleximers, inspired by styrene-based polymeric ionic liquids with fluorinated hydrophobic tails, required additional plasticization with an ionic liquid to lower the $T_g$ below the degradation temperature. Here, we prepare two new and improved versions, based on more flexible acrylate and methacrylate backbones and without perfluorinated groups (Figure \ref{ch4fig1}a). With these polymers we form two compleximers that differ solely in their backbone chemistry and that are both resistant to water and other solvents (Figure \ref{ch4_figA_5}). 
As shown in Figure \ref{ch4fig1}e and f, the compleximers can be reshaped easily at elevated temperatures below the degradation temperature (Figure \ref{ch4_figA_3}) in the absence of any plasticizer and without precise temperature control: upon heating a rectangular sample of methacrylate compleximer with a heat gun, the sample transforms from a hard glassy solid to a soft rubbery material, allowing us to twist it into a new shape (Movie \ref{ch4_movie3}). This behavior suggests a relatively broad thermal transition, reminiscent of the processability of silica glass and vitrimers. 

To  investigate this in more detail, we measure the dynamic elastic modulus of the materials as a function of temperature using dynamic mechanical analysis (DMA). 
We find that the glass transition is indeed very broad: for both compleximers, the storage modulus ($G’$) decreases gradually over a wide temperature range (Figure \ref{ch4fig1}g), approximately by an order of magnitude in a temperature window of 60 \textdegree C. This feature is accompanied by a broad peak in the loss modulus (Figure \ref{ch4fig1}h). The broadness of the transition makes it difficult to obtain a very accurate value for the glass transition temperature. For convenience, we define it here as the temperature corresponding to the maximum of the loss peak. A similar value for $T_g$ is found with differential scanning calorimetry (Figure \ref{ch4_figA_6}). Conventional thermoplastics have a much narrower glass transition: the same decrease in modulus for a thermoplastic polyvinyl acetate (pVAc) sample takes place over a temperature range of approximately 10 degrees \cite{wu2009correlations}. Interestingly, also for a strong glass former such as an epoxy vitrimer this decrease occurs over only 15 \textdegree C \cite{montarnal2011silica} (Figure \ref{ch4fig1}i). 
We also prepared a compleximer sample that was plasticized with 5 wt\% ionic liquid\cite{vanlange2024moderated}. As shown in Figure \ref{ch4fig1}g and h, this  leads to a decrease in the elastic modulus and a shift of the $T_g$ to lower temperatures, but a similarly broad transition.


\subsection*{Compleximers are strong glass formers}

\begin{figure}[t!]
\begin{center}
\includegraphics[width=\textwidth]{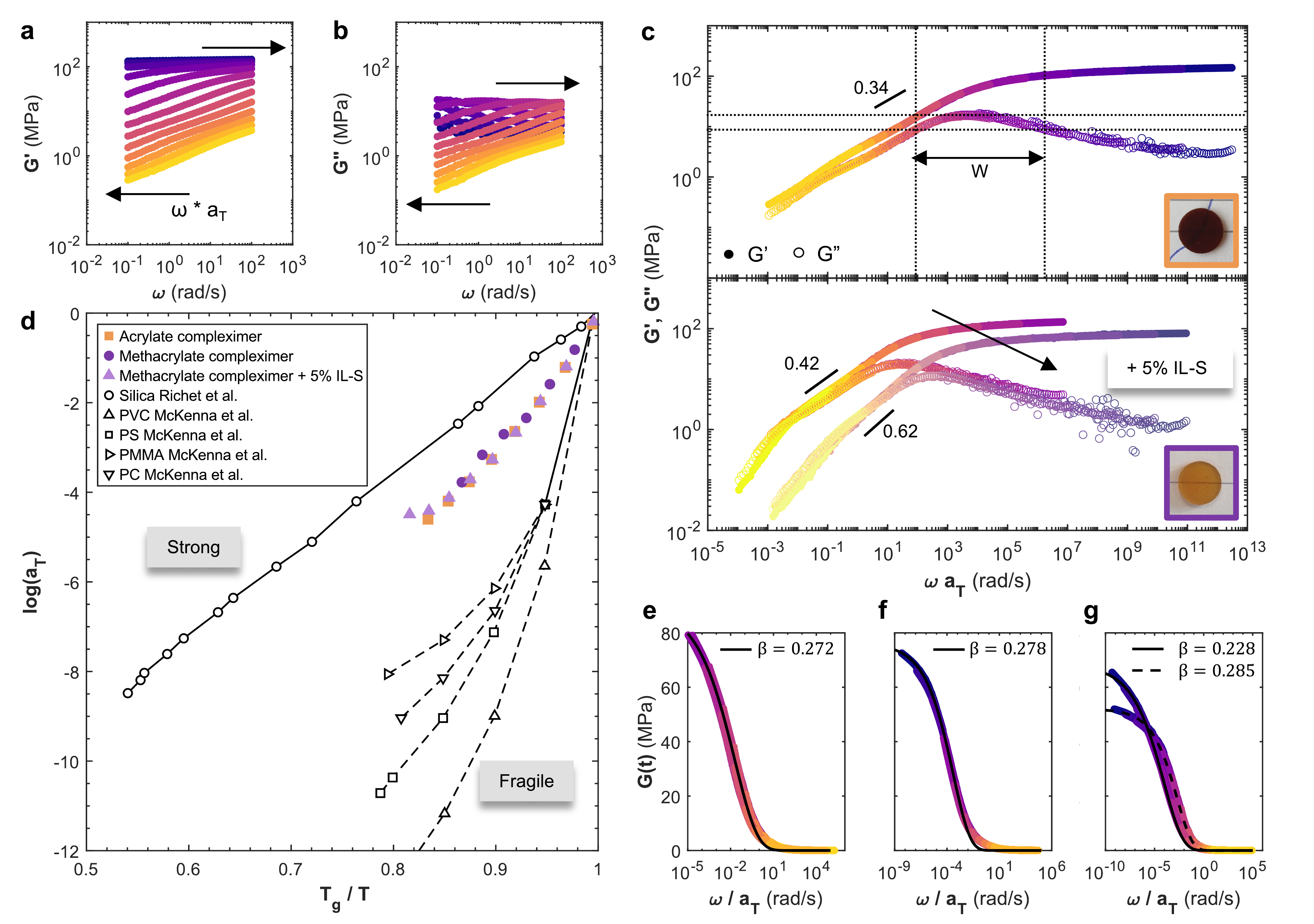}
\caption{\textbf{Compleximers have broad relaxation spectra and low fragility.} The frequency-dependent rheological measurements of the \textbf{(a)} storage modulus $G'$ and \textbf{(b)} loss modulus $G''$ of compleximers (here the acrylate compleximer is represented) can be shifted horizontally by multiplication of the frequency axis with a temperature-dependent shift factor $a_T$. \textbf{(c)} The time-temperature superposition (TTS) master curves of  acrylate compleximer (top), methacrylate (middle) and plasticized methacrylate (bottom) compleximers show broad power law regimes. The broadness of the relaxation spectra can be quantified through the full width at half maximum $W$ of the loss modulus. \textbf{(d)} Angell plot of compleximers. The black solid line corresponds to the Arrhenius law dependence of vitreous silica\cite{urbain1982viscosity}. Compleximers follow Arrhenius behavior, yet with an intermediate fragility index compared to other fragile thermoplastic melts\cite{huang2001new} represented with black dashed lines. The stress relaxation of compleximers (\textbf{(e)}: acrylate compleximer, \textbf{(f)}: methacrylate compleximer, \textbf{(g)}: methacrylate compleximer plasticized with 5\% ionic liquid, all for two different samples) is described by the Kohlrausch-Williams-Watts relation. The values for the stretch exponent $\beta$ are related to the width of the loss peak  as $\beta=1.14/W$. }
\label{ch4fig2}
\end{center}
\end{figure}

To investigate the relation between the  fragility of the glass transition and the viscoelastic relaxation spectrum of compleximers, we characterize the dynamic mechanical properties using the time-temperature superposition (TTS) principle, which allows us to probe rheological timescales that are otherwise inaccessible as a result of machine limitations. To this end, we measure the frequency-dependent storage and loss moduli ($G'$ and $G''$ respectively) at a range of temperatures (40-170 \textdegree C), which are then horizontally shifted on the frequency axis relative to a reference temperature by multiplying the frequencies with a temperature-dependent shift factor $a_T$ (Figure \ref{ch4_figA_7}, \ref{ch4_figA_8} and \ref{ch4_figA_9}). As shown in Figure \ref{ch4fig2}a-c, this leads to a master curve showing the viscoelastic signature of the materials over up to 16 decades in frequency. The fact that superposition works so well for these materials indicates that all relaxation processes in the material are affected in the same way by temperature. The acrylate and methacrylate compleximers show a very similar frequency dependence, with a plateau storage modulus around 200 MPa and a broad transition region at lower frequencies that follows a power law dependence over many decades in frequency. The gentle slope of the frequency sweeps implies that relaxation occurs over a wide range of frequencies, corresponding to a broad relaxation spectrum. The power law exponents 
vary somewhat between the different samples, and may be related to Rouse-like relaxation modes\cite{spruijt2013linear} or relaxation processes associated with the underlying network structure\cite{zaccone2014linking}.

The shift factors $a_T$ extracted from the TTS procedure are a measure for the underlying time scale of the relaxation processes. The decrease in the shift factors with increasing temperature thus indicates the enhanced dynamics at higher temperatures. To investigate whether compleximers are indeed strong glass formers, we construct a so-called Angell plot (Figure \ref{ch4fig2}d) by plotting the logarithm of the shift factor as a function of $T_g/T$, where we normalize the shift factors by their value at the glass transition \cite{angell1991relaxation, angell1995formation}. For comparison, we have also plotted the corresponding curves for a strong glass former (silica)\cite{urbain1982viscosity}, which follows the Arrhenius relation, and several fragile glass forming polymers\cite{huang2001new}, which can be described by the Williams-Landel-Ferry relation and show a much steeper dependence near $T_g$\cite{ljubic2014time}. This behavior near $T_g$ can be quantified by the fragility index, defined as the slope of the Angell plot at $T=T_g$ as described by Equation \ref{ch4eq1}. We find that all our compleximers have a fragility index of around 40 (Table \ref{ch4_figA__T1}), which is much lower than the fragility of  most polymers and closer to that of silica and vitrimers. We thus conclude that compleximers can be classified as (relatively) strong glass formers, in line with the broad thermal transition found in the mechanical response. To our knowledge, compleximers are thus the first organic strong glass formers stabilized by non-directional (ionic) interactions. 
We also note that the sample with the plasticizer shows a similar fragility as the unplasticized materials, which indicates that the plasticizer does not influence the nature of the relaxation process and its temperature dependence.

\subsection*{Compleximers have a broad relaxation spectrum}

Our dynamic measurements show that the relaxation spectrum of compleximers is broad, which means that the stress relaxation modulus deviates strongly from single-exponential (Maxwell-like) decay. Many glass formers show non-exponential decay, described by the stretched exponential function, Equation \ref{ch4eq2}. As shown by Boehmer et al., who collected data for a large number of different glass formers, there is a pronounced correlation between the stretch exponent $\beta$ and the fragility index $m$ of the same material, with strong glass formation (low $m$) corresponding to $\beta$ values close to 1\cite{boehmer1993nonexponential}. The relaxation of compleximers also follows a non-exponential decay (Figure \ref{ch4fig2}e-f and Figure \ref{ch4_figA_10}). Since not all glass formers are well described by a stretched exponential, it can be difficult to compare different materials. To get around this, we make use of the fact that the stretch exponent for a material that follows Equation \ref{ch4eq2} is directly related to the full width at half maximum $W$ of the peak in the loss modulus $G''(\log \omega)$, by $\beta\approx 1.14/W$ \cite{dixon1990specific}. We demonstrate the validity of this relation in Figure \ref{ch4fig2}e-f, where we fit the stress relaxation modulus of compleximers with Equation \ref{ch4eq2}, using the value of $\beta$ obtained through $W$ (Table \ref{ch4_figA__T1}) in Figure \ref{ch4fig2}c.



\begin{figure}[t!]
\begin{center}
\includegraphics[width=13cm]{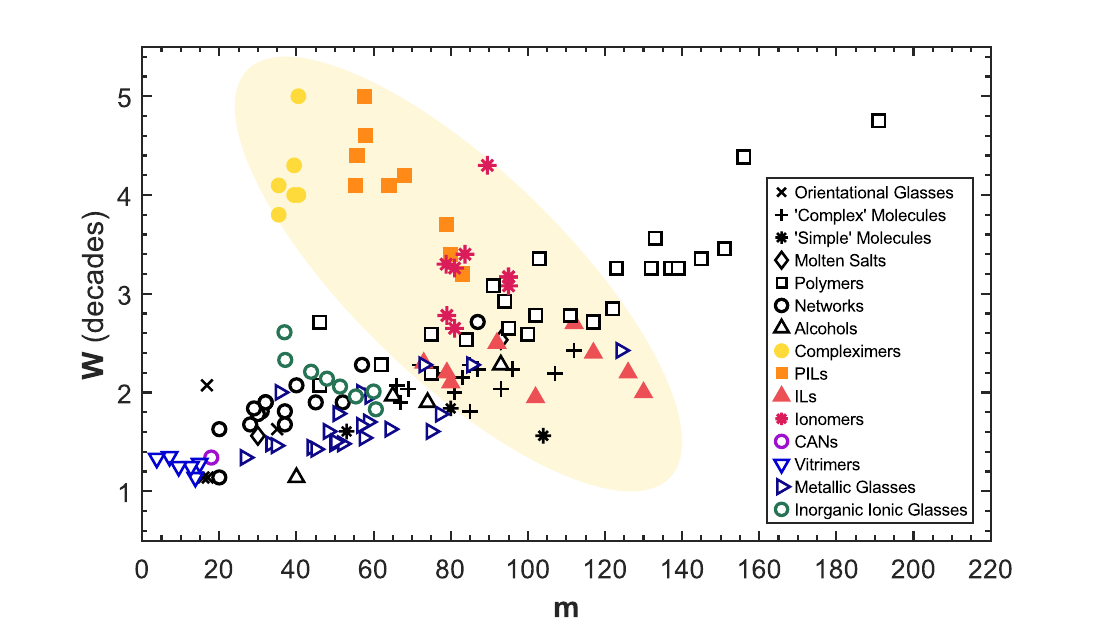}
\caption{\textbf{Ionic glass formers display a unique combination of low fragility and a broad relaxation spectrum} The width of the loss peak $W$ as a function of the fragility index $m$ shows a well documented, roughly linear trend for most glass formers, where the width of the relaxation spectrum increases with fragility\cite{boehmer1993nonexponential}. Compleximers, but also polymerized ionic liquids, ionomers, and ionic liquids follow an inverted trend, represented by the yellow shaded area. Green circles show data for an inorganic ionic glass former, Li\textsubscript{2}O/B\textsubscript{2}O\textsubscript{3}, \cite{matsuda2007calorimetric,kodama2002anharmonicity} which also follows an inverted trend, but with lower $W$.}
\label{ch4fig3}
\end{center}
\end{figure}

The relation between $\beta$ and $W$ allows us to use $W$ as a convenient parameter to compare different types of materials, irrespective of the nature of their relaxation. Figure \ref{ch4fig3} shows how the obtained width $W$ correlates with the fragility index for various glass formers as reported by Boehmer et al.\cite{boehmer1993nonexponential}. Here, we have also included literature data for various thermoplastics \cite{wu2009correlations}, vitrimers \cite{montarnal2011silica, meng2022rheology, elling2020reprocessable, roig2023disulfide, zhu2024catalyst, hong2024vitrimer, cywar2023elastomeric, denissen2015vinylogous}, dissociative covalent adaptable networks (CANs) \cite{elling2020reprocessable, rusayyis2021repcrocessable}, metallic glasses \cite{qiao2016dynamics} and three other types of polymers with ionic groups: ionomers \cite{chen2013ionomer}, polymers that are crosslinked by a small fraction of charged groups that form clusters with their counter-ions, ionic liquids (ILs) \cite{tao2015rheology} and polymeric ionic liquids (PILs)\cite{nakamura2013viscoelastic}, polyelectrolytes that feature an ionic liquid species in each monomer repeat unit (compleximers are essentially complexes formed by two oppositely charged polyionic liquids) (Table \ref{ch4_figA__T1} and \ref{ch4_figA__T2} and Note \ref{A.1}). All thermoplastics, vitrimers, CANs and metallic glasses fall on the same trend as the one reported previously\cite{boehmer1993nonexponential}, with low fragility correlating with a narrow relaxation spectrum. Our compleximers, however, deviate strongly from this trend, combining a broad relaxation spectrum with a low fragility. Interestingly, PILs also deviate from the reported trend, and while they are more fragile than compleximers, their relaxation spectrum is wider than would be expected based on the trend of Boehmer et al. Since PILs have a similar charge density as compleximers, this suggests that materials held together by ionic interactions follow a different trend than their uncharged equivalents. ILs and ionomers, with higher fragilities but narrower relaxation spectra, also follow this deviating trend for ionic glasses. As shown in Supplementary Note \ref{A.2}, this anomalous behaviour also explains the unusual broadness of the glass transition for compleximers shown in Figure \ref{ch4fig1}i.


\subsection*{Discussion}\label{sec12}
The opposite correlation between fragility and width of the relaxation spectrum found for charged organic materials suggests a special role of the long-ranged ionic interactions in these materials. 
To obtain further insight into the underlying mechanism responsible for this, we perform coarse-grained molecular dynamics simulations of model ionic liquids, polymerized ionic liquids, and compleximers (Figure \ref{ch4fig4}a). We perform simulations at constant pressure for a range of temperatures for each of these systems and measure the equilibrium density $\rho$ as a function of temperature (Figure \ref{ch4fig4}b). We find that compleximers have a higher density than PILs and ILs and a weaker temperature dependence of the density, i.e. a smaller thermal expansion coefficient $\alpha_T=1/\rho(d\rho/dT)_P$. The thermal expansion coefficient of a liquid depends on the interaction potential between its subunits, and to a reasonable approximation it varies with the potential depth $U_0$ as $\alpha_T\sim1/U_0$ \cite{lunkenheimer2023thermal, macdonald1955vibrational}. The small expansion coefficient of compleximers compared to ionic liquids can thus be explained by the covalent bonds between the monomers, which lead to a deeper attractive potential well (on average). We also find that the thermal expansion coefficient of compleximers is smaller than that of neutral polymers. This is a result of the  Coulomb interactions between the charged groups, which significantly increase the cohesive energy in the material. 

A recent theory argues that the fragility of a liquid increases proportionally to its thermal expansion coefficient, $m\sim\alpha_T$\cite{krausser2015interatomic}. This follows from the elastic shoving model, in which the energy barrier for relaxation $E_a$ is determined by the elastic work done in shoving aside the surrounding liquid\cite{dyre1998source, hecksher2015review}, which can be expressed as $E_a=V_cG_\infty$, where $V_c$ is the typical volume expansion required for a relaxation event and $G_\infty$ the high-frequency shear modulus of the material. Expansion of the material with increasing temperature leads to a decrease of the shear modulus $G_\infty$ and deviation from Arrhenius behavior. This provides a direct connection between the fragility and the thermal expansion coefficient and can explain why compleximers have a lower fragility than ionic liquids, and uncharged polymers. The origin of the broad relaxation spectrum of compleximers compared to ionic liquids can also be explained by their polymeric nature. As shown by computer simulations, covalent bonds in the polymer backbone give rise to an additional band in the vibrational density of states\cite{milkus2018interpretation}. This broadening of the vibration spectrum leads to additional relaxation modes and a lower stretch exponent $\beta$\cite{zaccone2020relaxation,cui2017relation}, as observed experimentally. This is also confirmed by our simulations, which show a lower stretch exponent for the $\alpha$-relaxation in the intermediate scattering function for compleximers than for PILs and ILs (Figure \ref{ch4fig4}c).

In conclusion, we have shown that organic materials with ionic interactions display an inverse correlation between fragility and the broadness of the relaxation spectrum, which sets them apart from other glass formers. This finding emphasizes the role of the nature of the cohesive interactions in the materials and may thus provide a new angle towards the development of a satisfactory theory of the glass transition. Moreover, our results suggest a way to develop novel materials that combine the processability of strong glass formers such as silica glass with the viscoelastic dissipation and damping properties of polymers.

\clearpage
\section{Methods}\label{sec11}

\subsection{Materials}

Acryloyl chloride (\( \geq \) 97\%), Azobisisobutyronitrile (AIBN, 98\%), Ammonium persulfate (APS, \( \geq \) 99\%), 3-Bromopropan-1-ol (97\%), Dichloromethane (DCM, anhydrous), Dimethylformamide (DMF, anhydrous), Magnesium sulphate (MgSO\textsubscript{4}, anhydrous), Methacryloyl chloride (\( \geq \) 97\%), 4-Methoxyphenol (99.8\%), Sodium chloride (\( \geq \) 99.5\%), 3-Sulfonylpropyl acrylate potassium salt (96\%), 3-Sulfopropyl methacrylate potassium salt (98\%), Tetrahydrofuran (THF, anhydrous), p-Toluenesulfonamide (98\%), Triethylamine (TEA, \( \geq \) 99\%), Trioctylamine (TOA, 98\%), Sodium bicarbonate (NaHCO\textsubscript{3}) and Ionic liquid methyl-trioctylammonium bis(trifluoromethylsulfonyl)imide (IL-S; $>$99\%) were obtained from Merck. Oxalyl chloride ($>$98\%) was purchased from TCI. Deuterated solvents were bought from Buchem. All other solvents were obtained from Biosolve.

\subsection{Synthesis of cationic polymers}
\subsubsection{Synthesis of 3-hydroxypropyl trioctylammonium bromide (HPTOA-Br)}

The synthetic procedure for the cationic polymer was adapted from Wrede et al. \cite{wrede2012polyelectrolyte}. Trioctylamine (6.36 g, 18.0 mmol) was added to a flask and degassed for 30 min with nitrogen before adding 3-bromo-1-propanol (10.0 g, 72.0 mmol) and degassing again for 10 min. The mixture was stirred at 60 \textdegree C for 2 days. The product was precipitated in 250 mL diethyl ether, filtered and washed with an additional 150 mL of ether. The product was dried under vacuum at 40\textdegree C overnight, yielding a white powder (6.49 g, 73\%). 

\subsubsection{Synthesis of 3-(Methacryloyloxy)propyl Trioctylammonium bromide (MAPTOA-Br) and 3-(Acryloyloxy)propyl Trioctylammonium bromide (APTOA-Br)}

HPTOA-Br (2.50 g, 5.07 mmol) was dissolved in 30 mL anhydrous DCM and degassed with nitrogen for 10 min before adding triethylamine (TEA, 2.06 g, 20.3 mmol). The solution was cooled to 0 \textdegree C and degassed again for 30 min, followed by the stepwise addition of methacryloyl chloride (1.60 g, 15.2 mmol) in 30 min. The reaction mixture was stirred overnight at RT. The excess methacryloyl chloride was quenched by adding 100 mL of a saturated aqueous NaHCO\textsubscript{3} solution to the obtained suspension, resulting in phase separation. The organic phase was washed 4 times with saturated NaHCO\textsubscript{3} solution and dried over MgSO\textsubscript{4}. The solvent was removed under vacuum at 25 \textdegree C, yielding MAPTOA-Br as a yellow oil (2.17 g, 3.87 mmol, 76\%).

To synthesize the acrylate monomer (APTOA-Br): acryloyl chloride (1.50 g, 16.4 mmol) was used instead of methacryloyl chloride and reacted with HP-TOA-Br (2.99 g, 6.10 mmol) and TEA (3.40 mL, 2.50 g, 24.4 mmol) according to the protocol above. This yielded in a clear, brown, viscous and sticky oil (2.66 g, 90\%).  

\subsubsection{Synthesis of poly(3-(Methacryloyloxy)propyl Trioctylammonium bromide) (pMAPTOA-Br) and Poly(3-(Acryloyloxy)propyl Trioctylammonium bromide) (pAPTOA-Br)}

To synthesize the polymethacrylate: MAPTOA-Br (2.97 g, 5.30 mmol) was dissolved in 30 mL anhydrous THF. The solution was degassed with nitrogen for 20 min and heated to 60 \textdegree C, before AIBN (8.69 mg, 0.0530 mmol) was added. The solution was degassed with nitrogen again for 10 min and then stirred at 60 \textdegree C overnight. The reaction mixture was concentrated under vacuum before precipitating in a cold mixture of 1:2 THF/H\textsubscript{2}O. The precipitate was dried under vacuum at 60 \textdegree C overnight, yielding a yellow, sticky polymer (1.33 g, 45\%), which was stored in a desiccator.

A few adaptions were made to the protocol for the polymerization of the acrylate. The polymerization was performed with APTOA-Br (1.46 g, 2.67 mmol) in 40 mL ethyl acetate and AIBN (22.9 mg, 0.13 mmol) under reflux conditions. The solvent was evaporated before the crude product was dissolved in THF and precipitation in cold water, to obtain a brown sticky solid (0.92 g, 63\%). 

\subsection{Synthesis of anionic polyelectrolytes}

\subsubsection{Synthesis of 3-(chlorosulfonyl)propyl methacrylate (SPMA-Cl) and 3-(chlorosulfonyl)propyl acrylate (SPA-Cl)}

The synthetic procedure was based on previous work\cite{vanlange2024moderated} and slightly adapted\cite{kammakakam2020tailored}. 3-Sulfopropyl methacrylate potassium salt (30.0 g, 122 mmol) was dissolved in about 50 mL anhydrous THF. The solution was degassed with nitrogen for 30 min and cooled to 0 \textdegree C before adding 3.4 mL of anhydrous DMF as a catalyst. The solution was degassed again, followed by the dropwise addition of oxalyl chloride (25.0 g, 197 mmol). The reaction was left overnight while stirring, allowing the ice bath to melt and the solution to rise to RT. Ice was added to the obtained suspension to quench the excess oxalyl chloride, resulting in phase separation. After decanting the upper aqueous layer, the organic layer was diluted with DCM. The organic solution was washed 6 times with water, then dried over MgSO\textsubscript{4}. The solvent was removed under vacuum at 25 \textdegree C, yielding SPMA-Cl as a light-yellow oil (18.3 g, 66\%).

For the synthesis of acrylate-based monomers 3-sulfopropyl acrylate potassium salt (14.59 g, 62.8 mmol) and 8.67 g (68.3 mmol) of oxalyl chloride were used with same molar ratios and purification method. SPA-Cl was obtained as a yellow oil (13.2 g, 89\%). 

\subsubsection{Synthesis of triethylammonium 1-[3-(methacryloyloxy)propylsulfonyl]-(p-toluenesulfonyl)imide (MATSI-TEA) and triethylammonium 1-[3(acryloyloxy)propylsulfonyl]-(p-toluenesulfonyl)imide (ATSI-TEA)}

p-Toluenesulfonamide (13.8 g, 80.5 mmol) was dissolved in 50 mL anhydrous THF and degassed with nitrogen for 30 min. The solution was cooled to 0 \textdegree C, before TEA (17.9 g, 177 mmol) was added, and the solution was degassed again. SPMA-Cl (18.3 g, 80.5 mmol) was dissolved in 30 mL anhydrous THF, degassed with nitrogen for 20 min and added to the reaction mixture. The reaction was stirred at 0 \textdegree C for 1 hour, then left stirring at room temperature overnight. The resulting suspension was filtered. A catalytic amount of 4-methoxyphenol was added to the filtrate as inhibitor in case of long-time storage of the monomer, before concentrating the solution under vacuum at 25 \textdegree C. The product was dissolved in DCM and the solution was washed 3 times with water. The remaining aqueous phases were extracted with ethyl acetate. Solvents were removed under vacuum at 25 \textdegree C, resulting in a yellow oil (MATSI-TEA, 9.57 g, 26\%).

To synthesize the acrylate-based monomer, the above synthesized SMA-Cl (11.9 g, 55.9 mmol, 1.06 Eq.) was used, with p-toluene sulfonamide (8.99 g, 52.5 mmol) and TEA (12.0 mL, 8.71 g, 86.1 mmol). The crude product was dissolved in DCM and extracted 3 times with water and once with brine. The water fractions were combined and washed three times with DCM. The combined DCM fractions are dried to obtain ATSI- TEA as a yellow oil (8.4 g, 37.5\%).

\subsubsection{Synthesis of poly(triethylammonium 1-[3-(methacryloyloxy)propylsulfonyl]-(p-toluenesulfonyl)imide) (pMATSI-TEA) and poly(triethylammonium 1-[3(Acryloyloxy)propyl sulfonyl]-(p-toluenesulphonyl)imide) (pATSI-TEA) }

MATSI was dissolved in 10 mL of THF, purified by aluminium oxide to remove the inhibitor and THF was removed under vacuum. MATSI (2.25 g, 4.86 mmol) was dissolved in 25 mL water. The solution was degassed with nitrogen for 20 min and heated to 60 \textdegree C, followed by the addition of APS (11.1 mg, 0.0482 mmol). The solution was degassed again for 10 min and then stirred at 60 \textdegree C overnight. The solvent was removed under vacuum. The product was dissolved in 6 mL dimethyl sulfoxide (DMSO) before precipitating in cold ethyl acetate. The precipitate was dried under vacuum at 60 \textdegree C overnight, yielding a yellow, sticky polymer (1.96 g, 87\%), which was stored in a desiccator.

For the polyacrylate: the above protocol was used with some small changes. The acrylate monomer ATSI (1.11 g, 2.46 mmol) was reacted with 1.0 mL of a degassed solution of APS (42.0 mg) was added. The reaction was heated to 90\textdegree C, resulting is a sticky dark, near black, product (0.59 g, 54\%).

\subsection{Nuclear magnetic resonance (NMR)}

\textsuperscript{1}H spectra were performed on a Bruker AV400 MHz spectrometer with CDCl\textsubscript{3}, CD\textsubscript{3}OD or D\textsubscript{2}O as a solvent. Results of all cationic and anionic compounds are shown in Figure \ref{ch4_figA_11}.

\subsection{Polyelectrolyte complexation}

The polycation and polyanion, methacrylate- or acrylate-based respectively, were each individually dissolved in a 1:1 mixture of acetonitrile/water at a concentration of 0.125 M. The solutions were mixed in equimolar amounts by adding them to an empty flask with a stirring bar, and were subsequently stirred overnight at room temperature. The resulting suspensions were filtered to remove the supernatant. The precipitates were resuspended in MilliQ water to wash out the small counterions under vigorous stirring. The water was replaced twice per day, which has been shown to remove counterions effectively\cite{RN38, vanlange2024moderated}, until the ionic conductivity of the supernatants reached equilibrium near the conductivity of MilliQ water. The complexes were isolated by filtration and dried in a vacuum oven at 60 \textdegree C at least overnight to obtain a white powder (1.40 g, 85\%) for the methacrylate compleximer, and a light brown powder (0.50 g, 43\%) for the acrylate compleximer. All compleximer powders were stored in a desiccator. 

\subsection{Conductivity measurements}

The conductivity of the supernatants from each compleximer washing cycle was measured using a Knick 703 conductometer equipped with a 2-electrode probe with a cell constant of 1.08 cm\textsuperscript{-1}.

\subsection{Compression-molding}

The compleximers were hot-pressed in a Specac manual hydraulic press equipped with heated plates to create circular (1 cm diameter) or rectangular (20 x 5 mm) samples using pressing dies obtained from Zhengzhou TCH Instrument Co. For the methacrylate compleximer, 100 mg or 130 mg compleximer powder was added to the circular pressing die or the rectangular pressing die respectively. The die was preheated for 15 minutes at 120 \textdegree C, before a pressure of 1.5 tons was applied for 15 minutes. For the acrylate compleximer, 90 mg was added to the circular pressing die, and the same procedure was executed at a processing temperature of 100 \textdegree C. For both compleximer types, after releasing the pressure, the die was removed from the press and cooled down for at least 30 minutes before the sample was removed. The systematic determination of the processing temperature involved gradually increasing the temperature until the sample was fully compact under the specified conditions.

\subsection{Plasticization with ionic liquid}

The methacrylate compleximer powder was plasticized with a screened ionic liquid, methyl-trioctylammonium bis(trifluoromethylsulfonyl)imide (IL-S). To achieve this, the compleximer powder was combined with 5\% of the total polymer mass of ionic liquid in a disposable 20 mL IKA ball milling tube. The powder and ionic liquid were milled with 15 stainless steel balls in three rounds of 3 minutes in the IKA ULTRA-TURRAX\textregistered Tube Drive at the highest setting. Between each round, the powder was scraped from the sides and the bottom of the tube for optimal mixing. After plasticization, the powder was hot-pressed as described before.

\subsection{Solubility of compleximers}

Hot-pressed methacrylate compleximer samples were immersed in a range of solvents (H\textsubscript{2}O, 2.5 M KCl in H\textsubscript{2}O, toluene, THF, acetonitrile and DMF) and kept for several weeks in closed containers to check their solubility.

\subsection{Moisture resistance}

The methacrylate compleximer sample that was immersed in water was patted dry with a paper towel, and measured after one day and after one week of immersion with extensional dynamic mechanical analysis. An amplitude sweep was performed at a frequency of 1 Hz to determine the plateau modulus at room temperature. This measurement was performed on an Anton Paar 702 Space Multidrive equipped with a DMA fixture.

\subsection{Thermogravimetric analysis (TGA)}

The thermal stability of the individual polymers and compleximer was determined using a Perkin Elmer STA 6000. Approximately 5 mg of sample was added to a ceramic sample cup and weighed. The sample was heated under an air flow of 20 mL min\textsuperscript{-1} at a rate of 10 \textdegree C min\textsuperscript{-1} from 30 to 600 \textdegree C and upon reaching 600 \textdegree C kept isothermal for 5 minutes. 
For the isothermal measurements, the samples were heated to the desired temperature at a rate of 50 \textdegree C min\textsuperscript{-1}. The methacrylate compleximer sample was kept isothermal for 30 minutes at 120 \textdegree C, 140 \textdegree C and 160 \textdegree C consecutively with intermediate heating rate of 20 \textdegree C min\textsuperscript{-1} to simulate degradation during material processing and mechanical testing. The acrylate compleximer was kept isothermal at temperatures between 100 \textdegree C and 160 \textdegree C with increments of 20 \textdegree C at a rate of 20 \textdegree C min\textsuperscript{-1}. All measurements were performed under an airflow of 20 mL min\textsuperscript{-1}. 

\subsection{Differential scanning calorimetry (DSC)}

We measured the enthalpic thermal transitions of our compleximers with a TA instruments DSC25, using aluminum Tzero pans. We performed temperature sweeps on approximately 7 mg of sample under a continuous nitrogen flow at 50.00 mL min\textsuperscript{-1}. An empty pan was used as reference. Each sample was equilibrated at -90 \textdegree C for 5 minutes, after which the sample was heated to 200 \textdegree C at a heating rate of 10 \textdegree C min\textsuperscript{-1}. The sample was kept isothermal at 200 \textdegree C for 5 minutes to ensure the completion of the heating ramp. The sample was cooled to -90 \textdegree C following the same procedure, and the entire cycle was repeated. Before doing the DSC cycles, separately prepared sample pans were equilibrated in an oven at 200 \textdegree C to confirm that the pans remain properly sealed during heating in the machine. The normalized heat flow was converted to the heat capacity through: 


\begin{equation}
C_p = \frac{\frac{dQ}{dt}}{M \cdot \frac{dT}{dt}}
\end{equation}

\noindent with $\frac{dQ}{dt}$ the heat flow, $M$ de mass of the sample and $\frac{dT}{dt}$ the heating rate.

\subsection{Dynamic mechanical analysis (DMA)}

The mechanical properties of the compleximers were analyzed using an Anton Paar 702 Space Multidrive. A DMA fixture was used for extensional oscillatory measurements on rectangular compleximer samples. Additionally, a 10 mm plate-plate geometry with a crosshatched surface on the upper plate was used for measurements in oscillatory shear mode on circular samples.\\
For both geometries, amplitude sweeps were performed on all samples to determine the linear viscoelastic regime, within which the amplitudes were chosen for subsequent measurements. Amplitude sweeps were performed at a frequency of 1 Hz. 

\subsubsection{Temperature sweeps}

Temperature sweeps of the methacrylate compleximers could only be completed over the entire temperature regime by splitting the sweep in two parts; by doing extensional tests at low temperatures, and tests in shear at high temperatures. This was because in shear, it was impossible to achieve sufficient contact between the material and the plate-plate geometry at low temperatures, and at high temperatures the sample would flow out of the extensional geometry. Therefore, in extension, temperature ramps were performed between 25 \textdegree C and 100 \textdegree C by applying a heating ramp followed by a cooling ramp (3 \textdegree C min\textsuperscript{-1}), while oscillating with a 0.01\% strain at a frequency of 1 Hz and with a preload force of 1500 Pa. In shear, temperature ramps were performed for temperatures from 90 \textdegree C up to 140 \textdegree C or 180 \textdegree C by applying a heating ramp followed by a cooling ramp (3 \textdegree C min\textsuperscript{-1}) while oscillating with a 0.01\% strain at a frequency of 1 Hz and with a Normal force of 2 N. Measurements were repeated twice in a row to ensure good contact between the sample and the plate. The extensional moduli were converted to shear moduli by dividing by 3 (assuming Poisson’s ratio to be 0.5). A small shift was observed between the shear and extensional data; to match the two curves, we multiplied the extensional moduli with 0.64. We deemed this reasonable, as the tan$\delta$ curves aligned without further shifting.\\
All temperature sweeps of the acrylate compleximers were performed in shear mode following the procedure as described above, but at a temperature range of 30-160 \textdegree C.

\subsubsection{Time-Temperature Superposition (TTS)}

Time-temperature superposition master curves were made from both frequency sweep data and stress relaxation data acquired at temperatures between 40 \textdegree C and 160 \textdegree C. All samples were tested in shear using a 10 mm crosshatched plate-plate geometry.\\

Frequency sweeps were performed in loops between 160 \textdegree C and 40 \textdegree C in decreasing steps of 10 \textdegree C. Samples were equilibrated at 170 \textdegree C for 5 minutes to ensure proper adhesion of the sample to the plates. In between cooling steps, the temperature was kept isothermal for 5 minutes to ensure proper equilibration. Frequency sweeps were performed from 100 to 0.01 rad s\textsuperscript{-1}, using a strain of 0.01\% and a Normal force of 2 N. 
Master curves were created by shifting the storage moduli data horizontally with respect to a reference temperature of 120 \textdegree C by multiplying the frequency axis of each temperature with a shift factor $a_T$. The same shift factors were used to shift the loss moduli and tan$\delta$. No vertical shift was applied.



Stress relaxation experiments were performed analogously to the frequency sweeps described above. After equilibration at the desired temperature, a strain of 0.01\% was applied, after which the stress relaxation was measured for 1000 seconds. Master curves were created by shifting the stress relaxation moduli horizontally with respect to the reference temperature of 120 \textdegree C.

\subsection{Molecular Dynamics simulations}
The molecular dynamics simulations were performed using a standard bead-spring model, with each bead having a size $\sigma$ and mass $m$. The polymers are modelled as branched polymers with backbone length $(N_{BB})$ $20$ , and each backbone monomer has a sidechain of length $(N_{SC})$ 1. All backbone monomers have a charge valency of $\pm 1$ and sidechain monomers possess no charge. The ionic liquid is modelled as a single charged bead $(N_{BB} =1 )$ connected to another neutral bead $(N_{SC} = 1 )$. All systems have a 1:1 stoichiometric charge ratio. The non-bonded interactions between the beads were modelled by Lennard-Jones interactions with strength $\epsilon_{LJ} = 1.0$ and cutoff $r_{cut} = 2.5$. The bonded interactions are modelled by a FENE potential with $k = 30$ and $R_0 = 1.5$. The charged beads also interact with a Coulomb potential $U_{coul} = q_i q_j/\epsilon r_{ij}$ where $q$ is the charge valency, $\epsilon$ is the dielectric constant and $r_{ij}$ is the distance between charges. The dielectric constant $\epsilon$ is set to 0.1 so as to achieve a strong enough electrostatic interaction strength with relaxation timescales that are reasonable for the simulations.\cite{yang2021influence} The MD simulations are performed in LAMMPS\cite{thompson_lammps_2022} and visualized with Ovito\cite{ovito}. 

The system is first equilibrated with a soft potential, followed by a small relaxation with electrostatic interactions disabled. Then, the strength of electrostatic interactions is gradually increased (to avoid abrupt changes in forces) by changing the dielectric constant from 0.8 to its final value of 0.1 over $500\tau$. The system is then equilibrated for $1.5 \times 10^4 \tau$, which is longer than the highest $\alpha$ relaxation time $\tau_\alpha$ across all systems simulated. $\tau_\alpha$ is measured as the time at which the self-intermediate scattering function, $F_s(q^*,t) = \langle \frac{1}{N} \sum_i^N e^{i\mathbf{q}^* \cdot [\mathbf{r_i}(t+t^{\prime}) - r_i(t^{\prime})]}\rangle$ where $\textbf{q}^*$ corresponds to the main peak of the static structure factor, $\mathbf{r_{i}}$ indicates the position of the \textit{i}th particle, and $\langle \cdot \cdot \cdot\rangle$ denotes an ensemble average, has a value of 0.2.  The production run is carried out for $3 \times 10^4 \tau$ in NPT ensemble using a Noose-Hoover thermostat at pressure $P = 0.1$. We choose this value of pressure because it is closer to the atmospheric pressure.\cite{liu2021effects}. Periodic boundary conditions are used during the simulation.







\clearpage





\backmatter

\bmhead{Supplementary information}

Supplementary information available







\subsection*{Acknowledgements}
This work was supported by the Dutch Research Council (NWO) OCENW.KLEIN.326 (J.v.d.G.).

\subsection*{Competing interests} 
The authors declare that they have no competing interests.

\subsection*{Data availability}
All data needed to evaluate the conclusions in the paper are present in the paper and/or the Supplementary Materials.

\subsection{Author contributions}
SvL, JS, and JvdG conceived the project. SvL, JvdG, DtB, AG, and NV developed the methodology. SvL, DtB, EB, JP, and MvN performed syntrhesis and the investigation. SvL, EB, MvN, and NV curated the data. SvL, EB, MvN, and JvdG carried out the data analysis. NV and AG performed the programming and code testing; SvL and JvdG validated modeling data with help of AG and AZ. SvL created the visualizations. SvL and JvdG wrote the original draft. All authors (SvL, DtB, EG, JP, MvN, NV, AZ, AG, JS, and JvdG) reviewed and edited the manuscript. JvdG and JS supervised the project. JvdG acquired the funding.


\clearpage

\counterwithin{figure}{section}

\begin{appendices}

\section{Supplementary Notes}\label{secA1}

\subsection{Determining the glass transition temperature, the fragility index,  and the width of the relaxation spectrum, listed in Table \ref{ch4_figA__T1} and \ref{ch4_figA__T2}}
\label{A.1}

\subsubsection{Angell plot}
The determination of the glass transition temperature using DMA and DSC is highly influenced by both frequency and heating rates. To standardize our approach, we define $T_g$ based on frequency sweep TTS master curves. These curves exhibit a distinct peak in the loss modulus ($G''$), one common indicator of the glass transition. The peak corresponding to the 'dynamic glass transition' is shifted to a reference frequency, here set to 1 Hz ($\omega a_T=2\pi$). We then renormalize all shift factors using this reference. Finally, $T_g$ is determined by interpolating to the point where the new shift factor equals 1.

\subsubsection{Fragility index $m$}

The fragility index is defined in Equation 1. It is thus determined by calculating the slope of the Angell plot of each material close to $T=T_g$. For some literature materials, $m$ was reported directly. For ionomers, Vogel Fulcher parameters $D$, $T_0$ and $T_g$ were reported. From those parameters, the fragility could be calculated through 

\begin{equation}
    m = D T_0 T_g/[(T_g -T_0)^2\ln10]
\end{equation}

\noindent with $D$ the dimensionless strength parameter and $T_0$ the Vogel temperature. Fragile glass formers show large deviations from Arrhenius behavior, corresponding to small values for $D$ ($D<10$)\cite{boehmer1993nonexponential}.



\subsubsection{The width of the relaxation spectrum represented by $\beta$ and $W$}

The width of the relaxation spectrum can be described by the full width at half maximum (in decades) of the peak in the loss modulus as a function of the shifted frequency $\omega a_T$ in the master curve \cite{boehmer1993nonexponential}. When available, $W$ was directly calculated from the frequency sweep master curves. For materials for which only stress relaxation curves were reported, we fitted the curve with a stretched exponential 
$G(t) = G_0 \exp\!\left( -\left(\tfrac{t}{\tau}\right)^{\beta} \right)$  from which we extracted the stretching exponent $\beta$. We then obtain the width using  $\beta \approx 1.14/W$. We verify this relationship for our compleximers by plotting the stretched exponential function obtained in this way over our stress relaxation experiments (Figure 2e-g). We indeed find that the stretch factors found through the full width half maximum method describe the curves well.

\subsection{The steepness of the glass transition}
\label{A.2}
As shown in Figure 1i, compleximers have a very broad glass transition. This broadness can be characterized by the slope of the curve showing $\log G'$ (measured at a frequency of 1 Hz) as a function of temperature at the steepest point in the thermal transition. We normalize this slope by the glass transition temperature to obtain a dimensionless steepness parameter $S$, defined as:
\begin{equation*}
    S=-T_g \left.\frac{d \log_{10} G'}{dT}\right|_{T_g}
\end{equation*}

In figure \ref{ch4fig3Steepness}a, we show the steepness as a function of the fragility index for our compleximers, as well as for a range of other materials.
For some materials, temperature sweeps were unavailable. For ionomers\cite{chen2013ionomer}, d$\log E'/$d$(T_g/T)$ was reported. We assume this is approximately the same as $-T_g*$d$\log G'/$d$T$. For PILs\cite{nakamura2013viscoelastic}, the slopes were obtained by combining the TTS frequency sweeps with the shift factors.

As can be seen in Fig. \ref{ch4fig3Steepness}a, compleximers have a very low steepness index (corresponding to a very broad thermal transition). While there is an overall
increasing trend of the steepness with increasing fragility, the ILs, PILs, ionomers, and
compleximers  appear to follow a different trend than the other glass formers, as was the case for the relation between $W$ and $m$. 
In
particular, compleximers have a much lower steepness than, for example, vitrimers while the latter have a lower fragility. We hypothesize that this is related to the broadness of
the relaxation spectrum. We show this by breaking down the steepness parameter as

\begin{equation*}
    S=-T_g \frac{d \log_{10} G'}{dT}=-T_g \frac{d \log_{10} G'}{d\log_{10} (\omega a_T)}\frac{d\log_{10} (\omega a_T)}{dT}\approx n \cdot m
\end{equation*}
where $d \log_{10} G'/d\log_{10} (\omega a_T)$ is the slope of the TTS master curve of the storage modulus, which we denote as $n$, and where $-T_gd\log_{10} (\omega a_T)/dT$ corresponds to the fragility index $m$.

Compleximers have a relatively low power law exponent $n\approx0.4$. By comparison, vitrimers show an almost single-exponential relaxation time Maxwell behaviour, which corresponds to $n\approx2$, explaining why they have a relatively steep transition despite their low fragility. In Note \ref{A.3} we show that, to good approximation, $n\sim\beta\sim1/W$, suggesting that $S\sim m/W$. Indeed, as shown in Figure \ref{ch4fig3Steepness}, we obtain a better data collapse of all material types by plotting the steepness $S$ as a function of $m/W$.










\subsection{Frequency sweep for the Kohlrausch-Williams-Watts relation}
\label{A.3}

In Note \ref{A.2} it was shown that the steepness of the modulus variation with temperature around $T_g$ can be expressed as
\begin{equation*}
-T_g\frac{d\log_{10}G'}{dT}\approx m \cdot n
\end{equation*}
with $m$ the fragility index and $n$ the slope of a plot of  $\log_{10}G'$ versus $\log_{10}\omega$. For most materials plotted in Figure \ref{ch4fig3Steepness}, only the stretch exponents $\beta$ of the Kohlrausch-Williams-Watts (KWW) relaxation function are reported in literature. To estimate the exponent $n$, we need to convert the KWW relaxation function to the frequency domain. We do this numerically, by first writing 
\begin{equation*}
G(t)=G_0e^{-\left(t/\tau\right)^\beta}=G_0\int_0^\infty e^{-t/u}\rho(u)du
\end{equation*}
where $\rho(u)$ is the relaxation spectrum corresponding to the KWW relaxation function, which can be written as\cite{lindsey1980detailed}:
\begin{equation*}
\rho(u)=-\frac{1}{\pi u}\sum_{k=0}^\infty \frac{(-1)^k}{k!}\sin(\pi\beta k) \Gamma(\beta k+1)\left(\frac{u}{\tau}\right)^{\beta k}
\end{equation*}
where $\Gamma(x)$ denotes the Gamma function. The storage modulus $G'$ in the frequency domain can then be written as
\begin{equation*}
G'(\omega)=G_0\int_0^\infty\frac{(\omega u)^2}{1+(\omega u)^2}\rho(u)du
\end{equation*}

\begin{figure}[h!]
\begin{center}
\includegraphics[width=200pt]{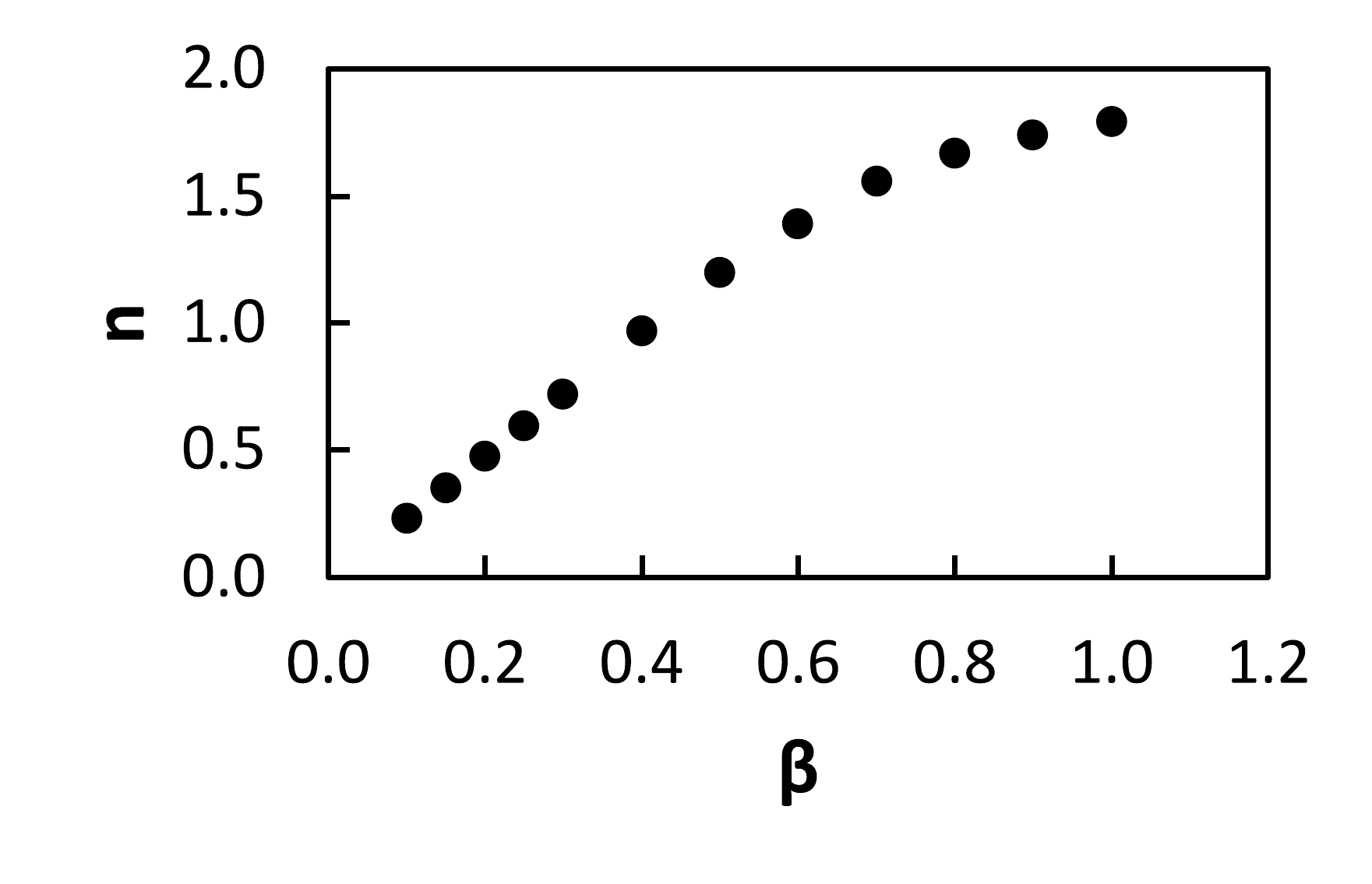}
\caption{\textbf{Log-log slope of the storage modulus versus frequency, $n$ as a function of the stretch exponent $\beta$ for a material with a stretched exponential stress relaxation modulus.}}
\label{nvsbeta}
\end{center}
\end{figure}

We then calculate the relaxation spectrum $\rho(u)$ numerically for different values of $\beta$, and use this to calculate the corresponding frequency-dependent moduli. Since the slope of the frequency sweeps varies with frequency, we estimate $n$ as the slope at the frequency where $G'(\omega)=G_0/10$. In figure \ref{nvsbeta}, we show the resulting $n$-values as a function of $\beta$. It can be seen that lower stretch exponents correspond to lower $n$-values, and for $\beta$ smaller than roughly 0.7, we have approximately $n\sim\beta\sim1/W$. 

\clearpage

\section{Extended Data}



\begin{figure}[h!]
\begin{center}
\includegraphics[width=\textwidth]{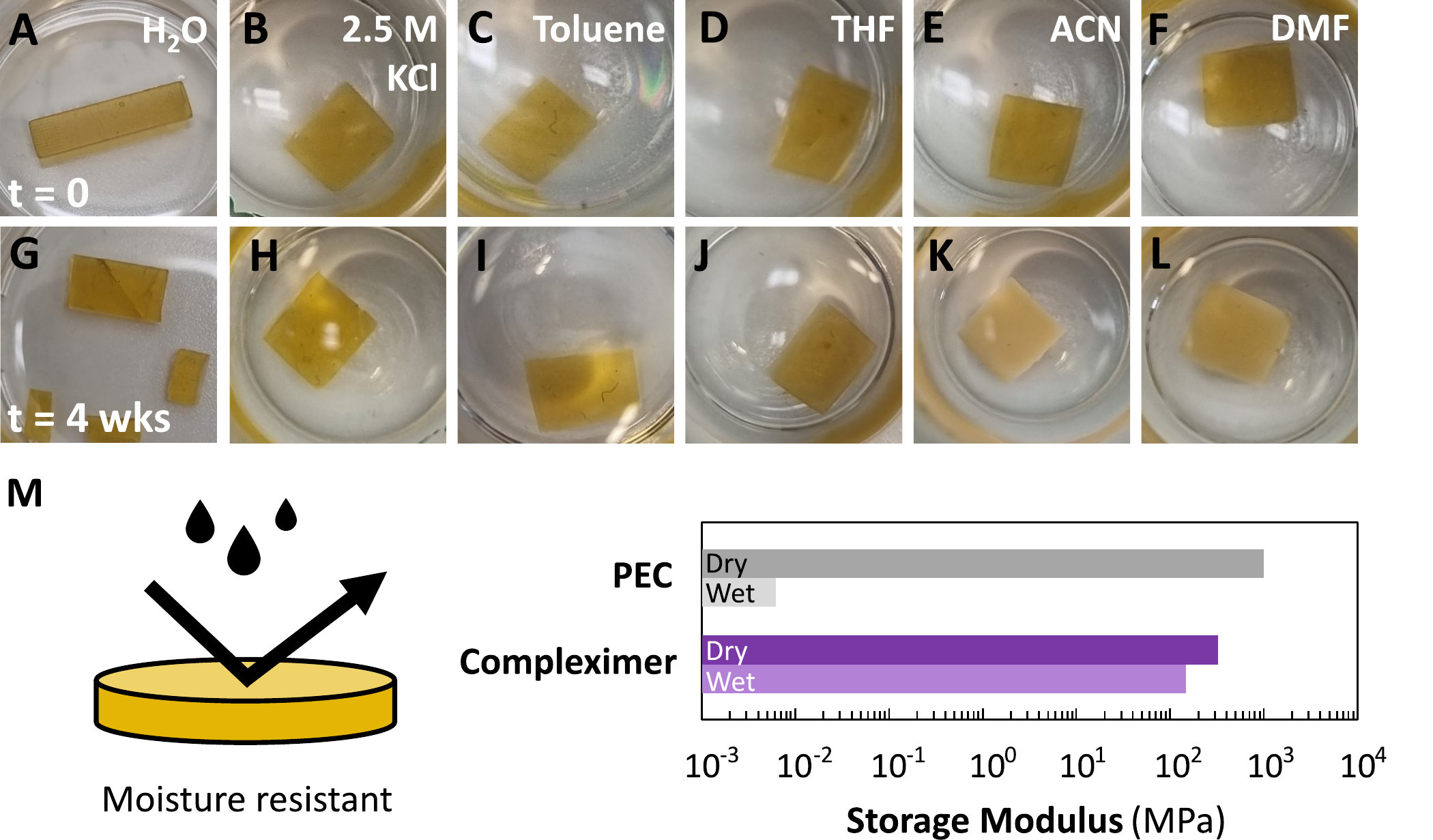}
\caption{\textbf{The solubility and water-resistance of methacrylate compleximer.} Compleximer samples directly after immersion in \textbf{(A),} water, \textbf{(B)}, 2.5 M KCl in water, \textbf{(C)}, toluene, \textbf{(D)}, THF, \textbf{(E)}, ACN and \textbf{(F)}, DMF. After four weeks of immersion the compleximer samples in \textbf{(G)}, water, \textbf{(H)}, 2.5 M KCl, \textbf{(I)}, toluene and \textbf{(J)}, THF have not visibly changed. The sample in water was broken during a DMA measurement. Samples in \textbf{(K)}, ACN and \textbf{(L)}, DMF have turned white and opaque, which suggest the uptake of some solvent. No visual dissolution has occurred. \textbf{M}, Compleximers show good resistance to water, unlike hydrophilic polyelectrolyte complexes (PECs).}
\label{ch4_figA_5}
\end{center}
\end{figure}

\begin{figure}[h!]
\begin{center}
\includegraphics[width=\textwidth]{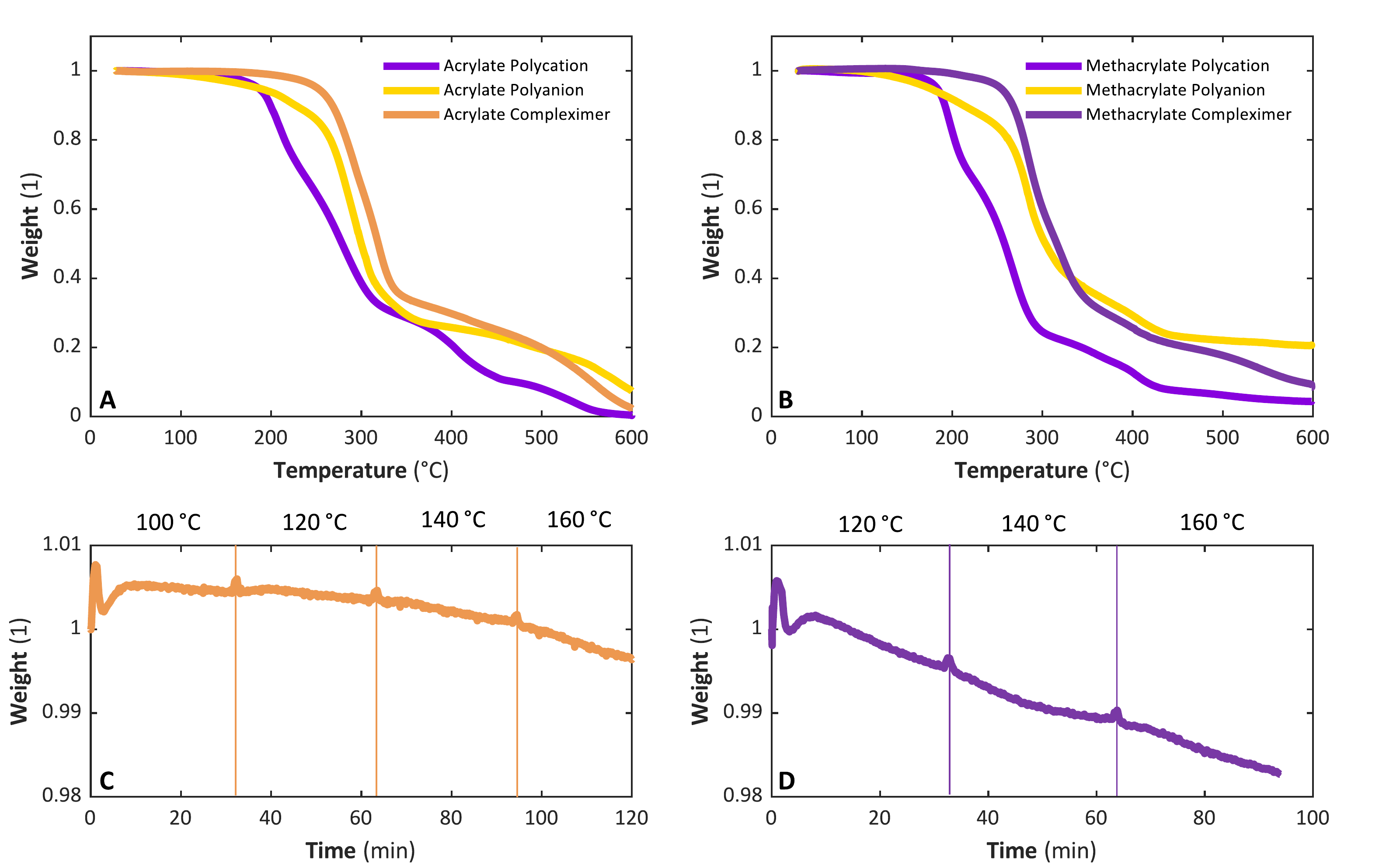}
\caption{\textbf{Thermogravimetric analysis (TGA) of compleximers:} degradation profiles of \textbf{(A)}, acrylate and \textbf{(B)}, methacrylate compleximers show that the complex has higher thermal stability than the individual polyelectrolytes from which they are composed. Isothermal TGA profiles over extended time indicate only minor degradation ($<$1\%) of the \textbf{(C)}, acrylate compleximer over a range of temperatures at and above the processing temperature. The methacrylate compleximer \textbf{(D)} shows some slight thermal degradation, approximately 0.5\% per half hour of heating above the processing temperature, which may result in some minor degradation during processing and characterization at elevated temperatures.}
\label{ch4_figA_3}
\end{center}
\end{figure}



\begin{figure}[h!]
\begin{center}
\includegraphics[width=200pt]{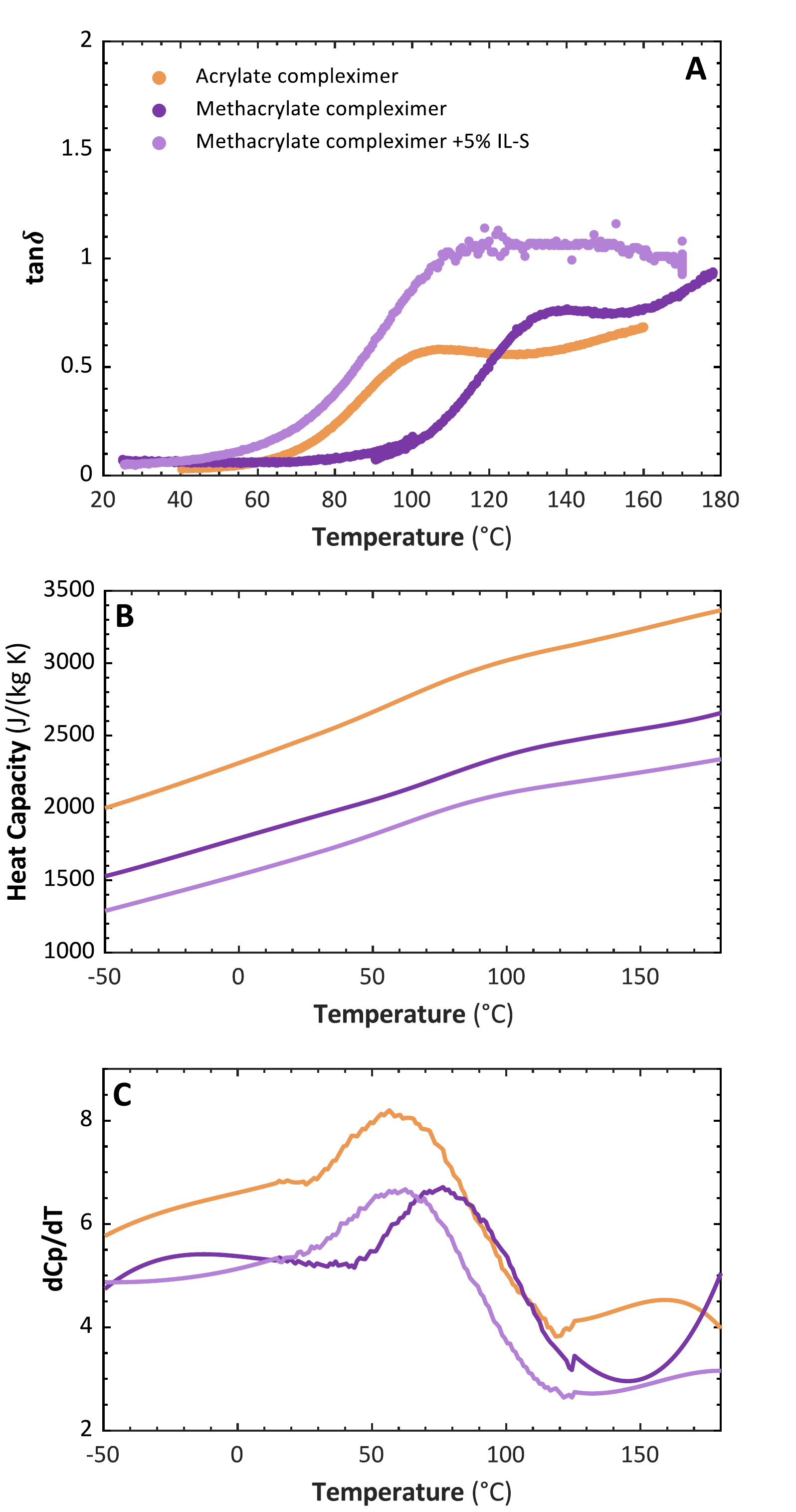}
\caption{\textbf{The broad thermal transition of compleximers measured with dynamic mechanical analysis and differential scanning calorimetry.} \textbf{(A)} The tan$\delta$ values of the temperature sweeps of compleximers. A broad transition is observed, represented by a wide peak. \textbf{(B)} The heat capacity as a function of temperature measured with DSC shows a very subtle shift in slope around the thermal transition. \textbf{(C)} The derivative of the heat capacity shows a more defined peak in the thermal transition. The thermal transitions measured with DMA and DSC are not at exactly the same temperature, this may be a result of the employed heating rates.}
\label{ch4_figA_6}
\end{center}
\end{figure}

\begin{figure}[h!]
\begin{center}
\includegraphics[width=\textwidth]{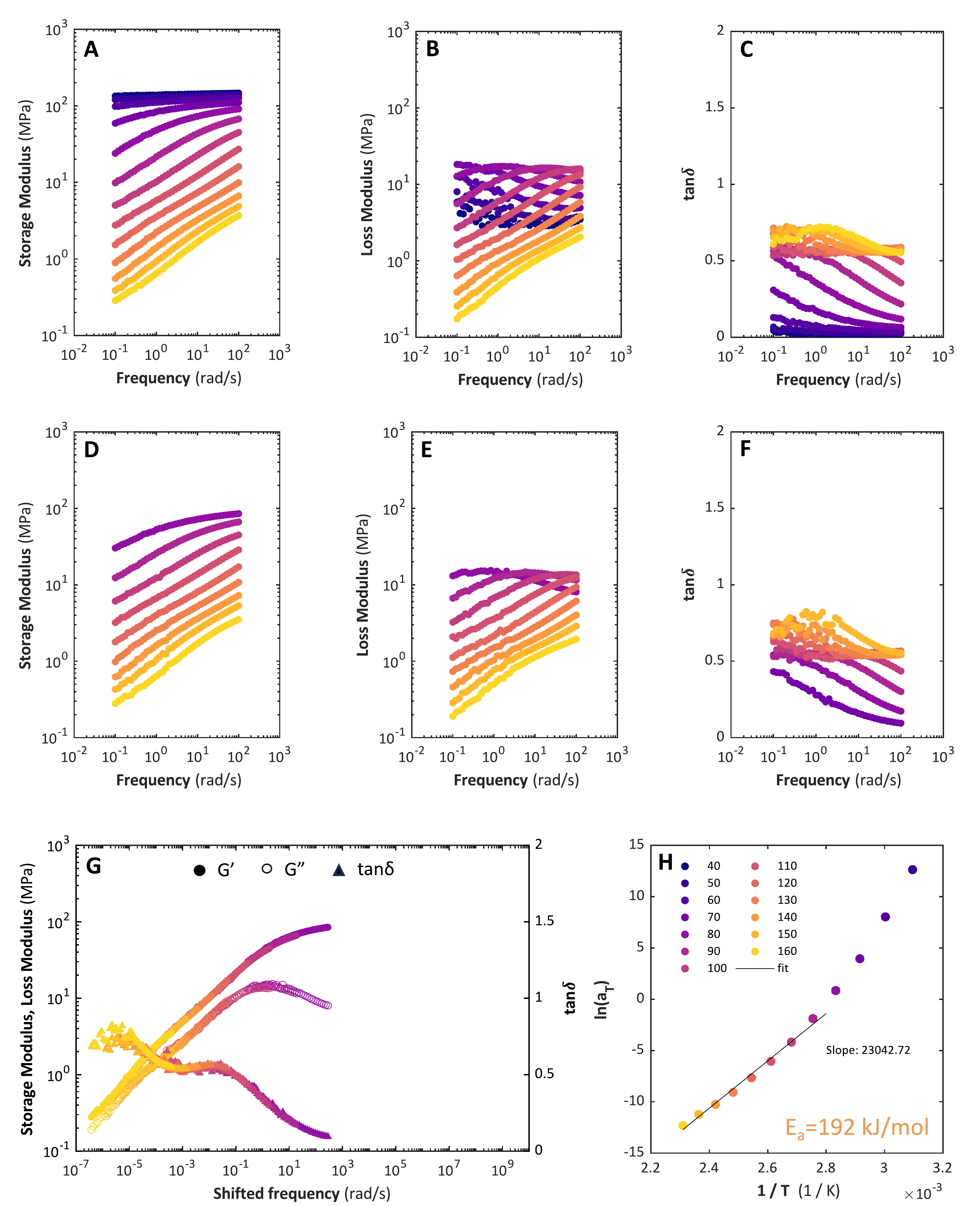}
\caption{\textbf{Time-temperature superposition data of acrylate compleximer.} \textbf{(A)} Storage modulus, \textbf{(B)} Loss modulus and \textbf{(C)} tan$\delta$ of one sample. \textbf{(D)} Storage modulus, \textbf{(E)} Loss modulus and \textbf{(F)} tan$\delta$ of a second compleximer sample. \textbf{(G)} Master curve of the second sample generated through shifting the data horizontally. \textbf{(H)} Natural logarithm of the shift factor as a function of the inverse temperature. An Arrhenius law is fitted through the data above $T_g$, the activation energy $E_a$ is determined from the slope.}
\label{ch4_figA_7}
\end{center}
\end{figure}

\begin{figure}[h!]
\begin{center}
\includegraphics[width=\textwidth]{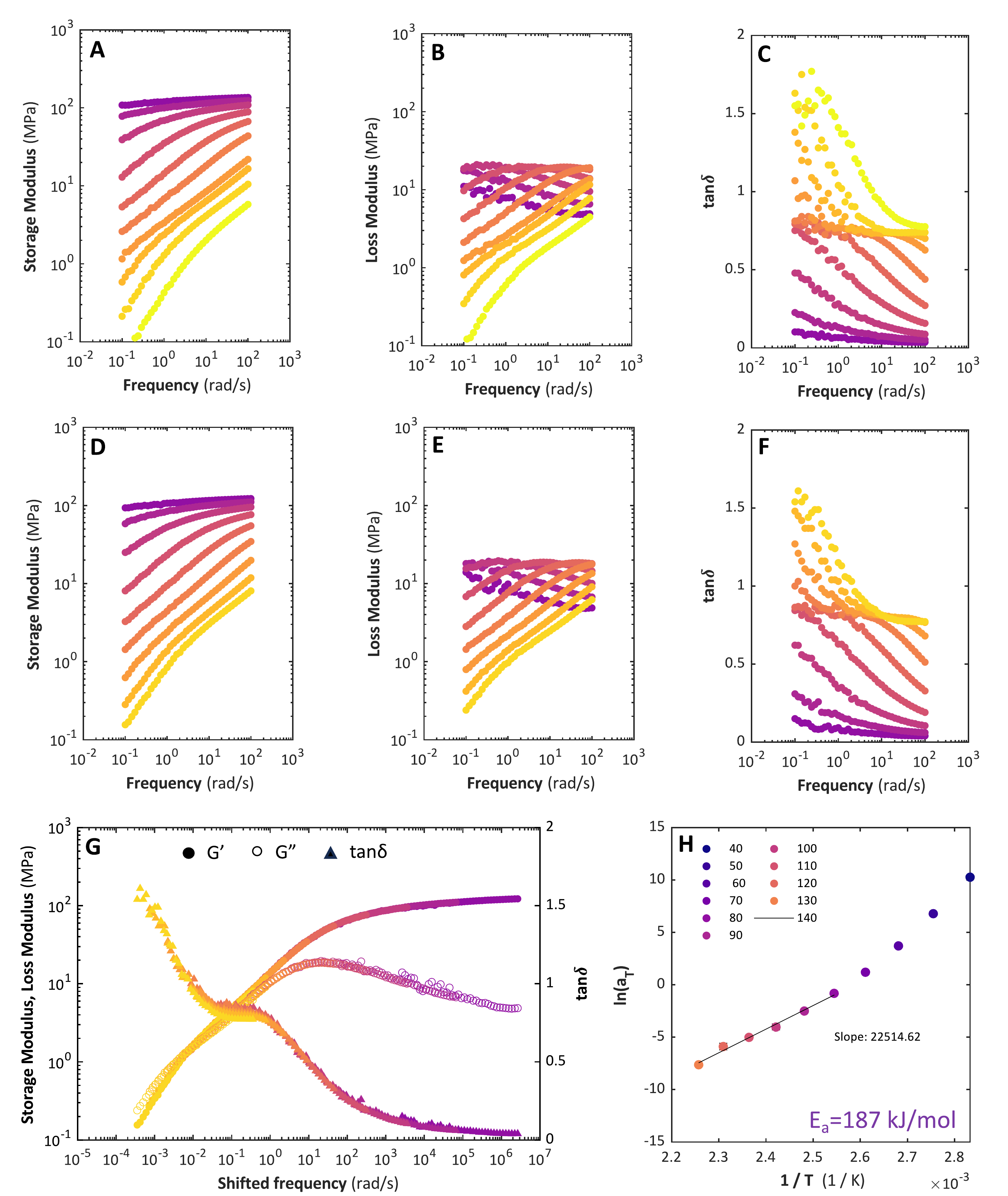}
\caption{\textbf{Time-temperature superposition data of methacrylate compleximer.} \textbf{(A)} Storage modulus, \textbf{(B)} Loss modulus and \textbf{(C)} tan$\delta$ of one sample. \textbf{(D)} Storage modulus, \textbf{(E)} Loss modulus and \textbf{(F)} tan$\delta$ of a second compleximer sample. \textbf{(G)} Master curve of the second sample generated through shifting the data horizontally. \textbf{(H)} Natural logarithm of the shift factor as a function of the inverse temperature. An Arrhenius law is fitted through the data above $T_g$, the activation energy $E_a$ is determined from the slope.}
\label{ch4_figA_8}
\end{center}
\end{figure}

\begin{figure}[h!]
\begin{center}
\includegraphics[width=\textwidth]{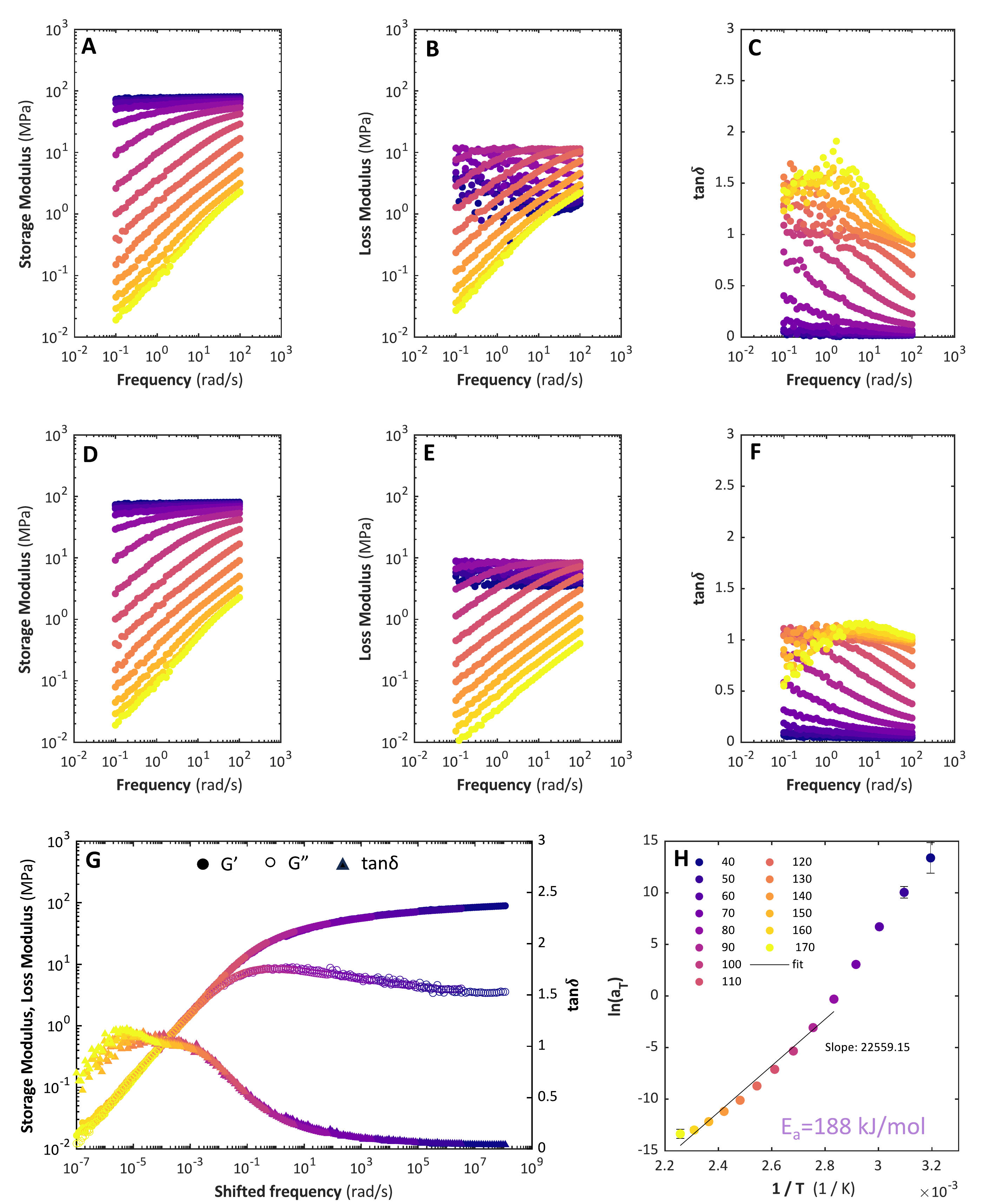}
\caption{\textbf{Time-temperature superposition data of methacrylate compleximer plasticized with 5\% IL-S.} \textbf{(A)} Storage modulus, \textbf{(B)} Loss modulus and \textbf{(C)} tan$\delta$ of one sample. \textbf{(D)} Storage modulus, \textbf{(E)} Loss modulus and \textbf{(F)} tan$\delta$ of a second compleximer sample. \textbf{(G)} Master curve of the second sample generated through shifting the data horizontally. \textbf{(H)} Natural logarithm of the shift factor as a function of the inverse temperature. An Arrhenius law is fitted through the data above $T_g$, the activation energy E\textsubscript{a} is determined from the slope.}
\label{ch4_figA_9}
\end{center}
\end{figure}

\begin{figure}[h!]
\begin{center}
\includegraphics[width=\textwidth]{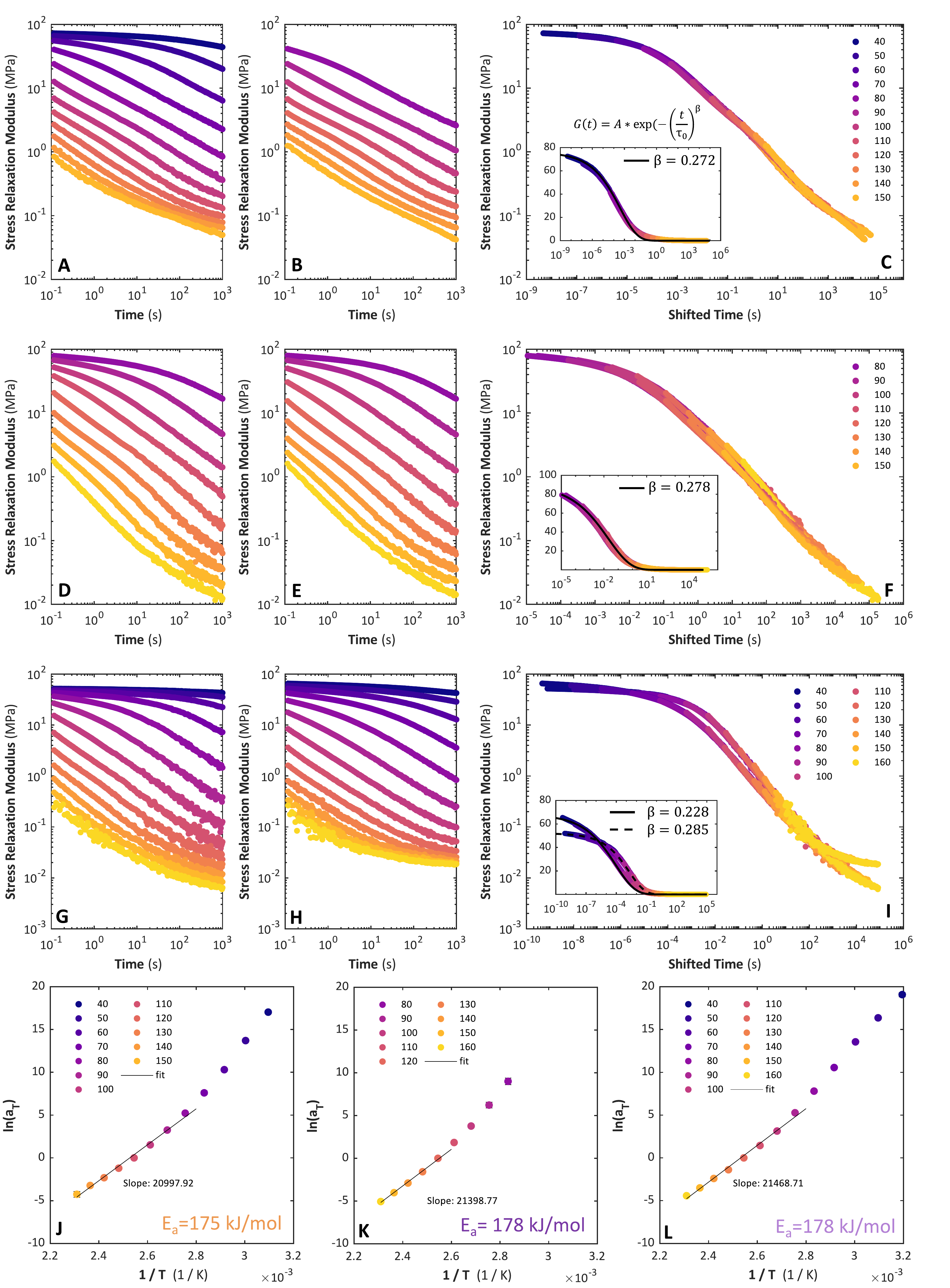}
\caption{\textbf{Stress relaxation of compleximers.} \textbf{(A,B)} Stress relaxation curves at several temperatures of two acrylate samples. \textbf{(C)} Shifting the data with a temperature dependent shift factor (by doing $a_T/t$) yields master curves. The inset displays the fitting of the mastercurve with a stretched exponential. The stretch factor $\beta$ is here the same as the width of the relaxation spectrum via $\beta$=1.14/W. The same is depicted for \textbf{(D,E,F)} two methacrylate samples. And \textbf{(G,H,I)} two methacrylate samples plasticized with 5\% IL-S. In \textbf{(J,K, and L)} the temperature dependent shift factors $a_T$ are plotted as a function of temperature. Above $T_g$, the data was fitted to yield the activation energy following Arrhenius Law. The found activation energies are very close to those found with the frequency sweep experiments.}
\label{ch4_figA_10}
\end{center}
\end{figure}

\begin{figure}[h!]
\begin{center}
\includegraphics[width=\textwidth]{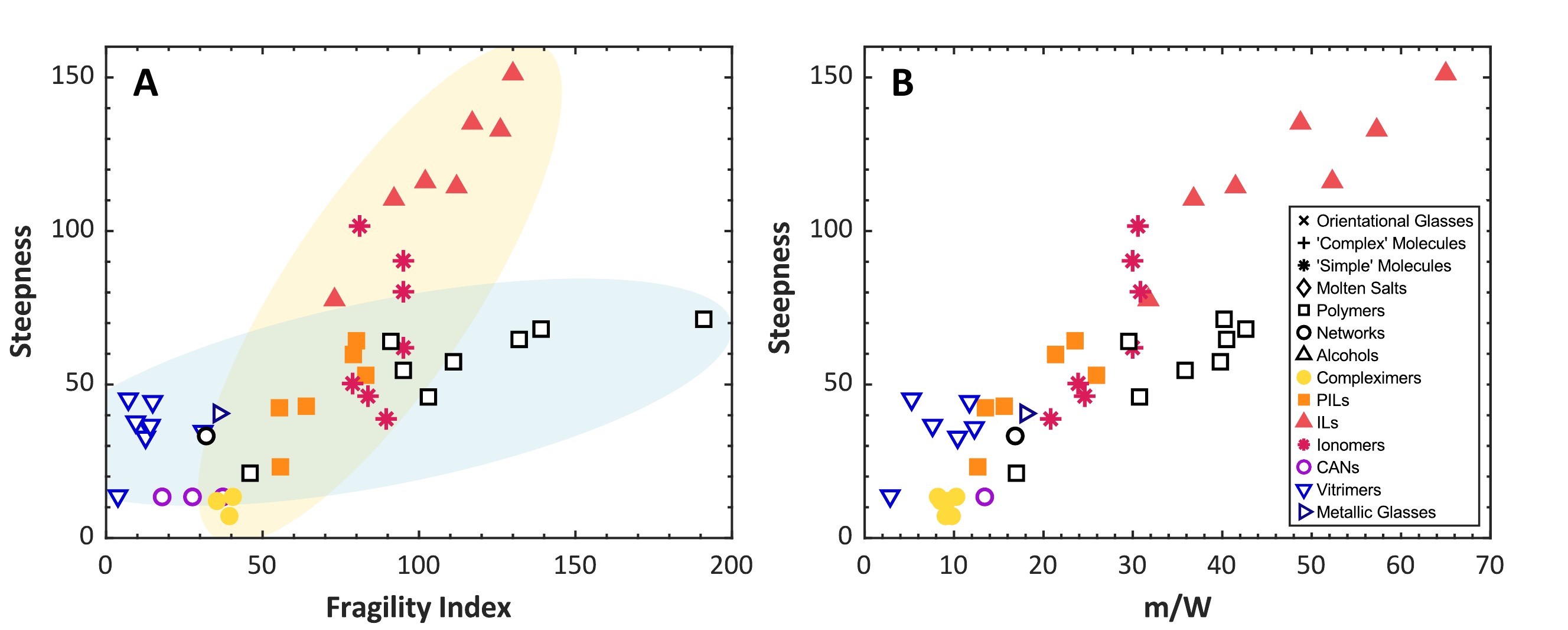}
\caption{\textbf{The steepness of the glass transition depends on the fragility and the broadness of the relaxation spectrum} \textbf{(A)}, The steepness parameter, $S=-T_gd\log _{10}G'/dT$ at 1 Hz in the steepest point of the temperature ramp plotted as a function of the fragility index. \textbf{(B)}, We collapse all data reasonably well by dividing the fragility index $m$ by the full width at half maximum of the loss peak, $W$.}
\label{ch4fig3Steepness}
\end{center}
\end{figure}

\begin{figure}[h!]
\begin{center}
\includegraphics[width=\textwidth]{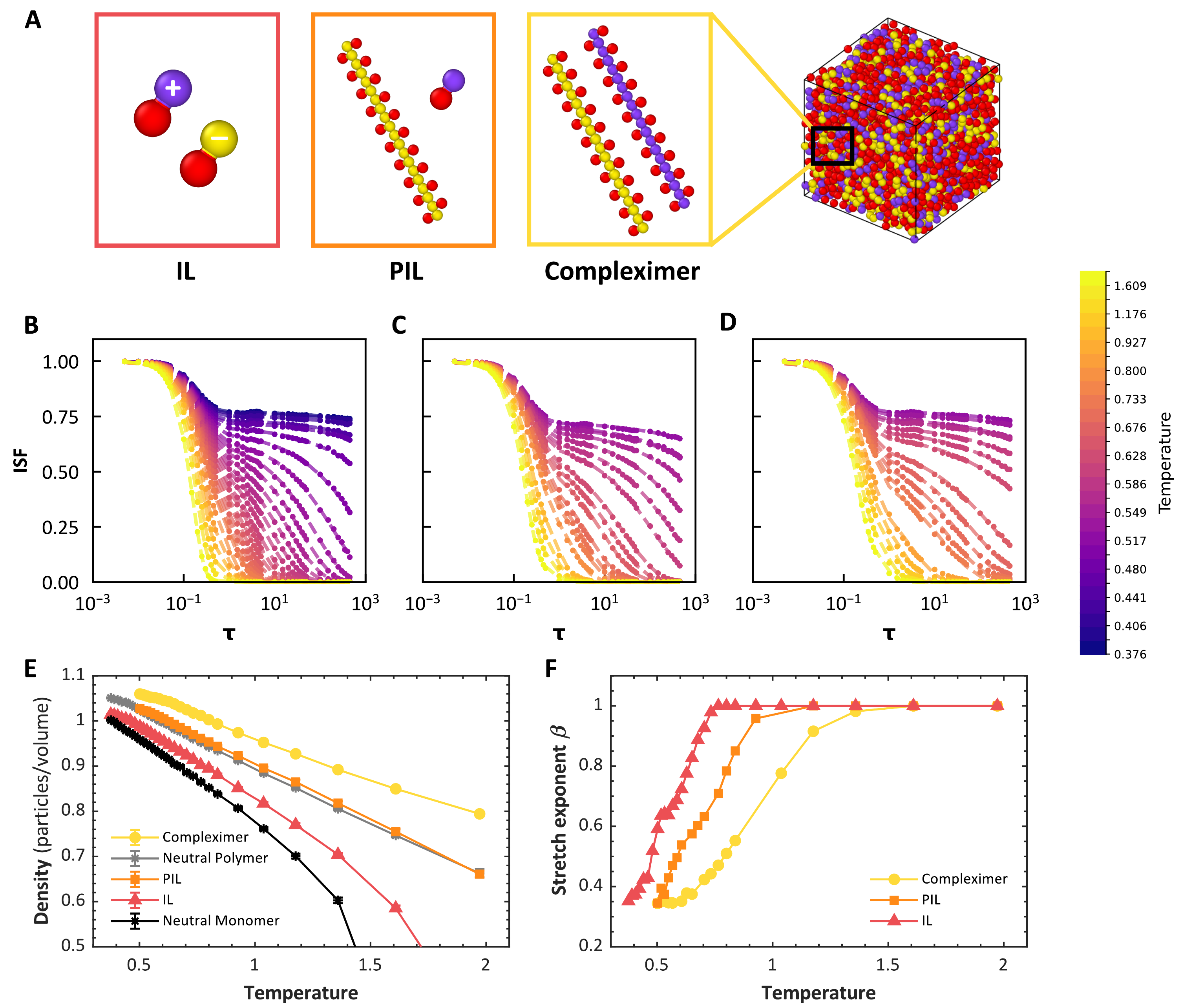}
\caption{\textbf{The effect of ionic bonds and connectivity on the density, expansion coefficient and stretch exponent.} \textbf{(A)}, We extract the density of various systems from Molecular Dynamics (MD) simulations. We use a coarse-grained bead-spring model with Coulomb interactions. An ionic liquid (IL) consists of charged beads (yellow and purple) with a neutral bead (red) attached. A polymerized ionic liquid (PIL) combines a polymerized IL with an oppositely charged monomeric IL. A compleximer contains both oppositely charged polymerized ILs with $N_{backbone}=20$ beads. \textbf{(B-D)}, The intermediate scattering functions (ISFs) are measured at a range of temperatures for IL \textbf{(B)}, PIL \textbf{(C)} and Compleximer \textbf{(D)} respectively. \textbf{(E)}, Ionic bonds lead to a higher density and a smaller expansion coefficient at $T_g/T\approx 0.9$ ($\alpha_T=-0.23$) than for neutral polymers ($\alpha_T=-0.28$). Compleximers have lower expansion coefficient than ILs ($\alpha=-0.32$) and similar to PILs ($\alpha_T=-0.23$). ILs have a higher density than their neutral equivalents. \textbf{(F)}, The stretch exponent $\beta$ of the $\alpha$-relaxation, found by fitting the intermediate scattering functions, is lower for compleximers than for PILs and ILs.}
\label{ch4fig4}
\end{center}
\end{figure}

\begin{figure}[h!]
\begin{center}
\includegraphics[width=\textwidth]{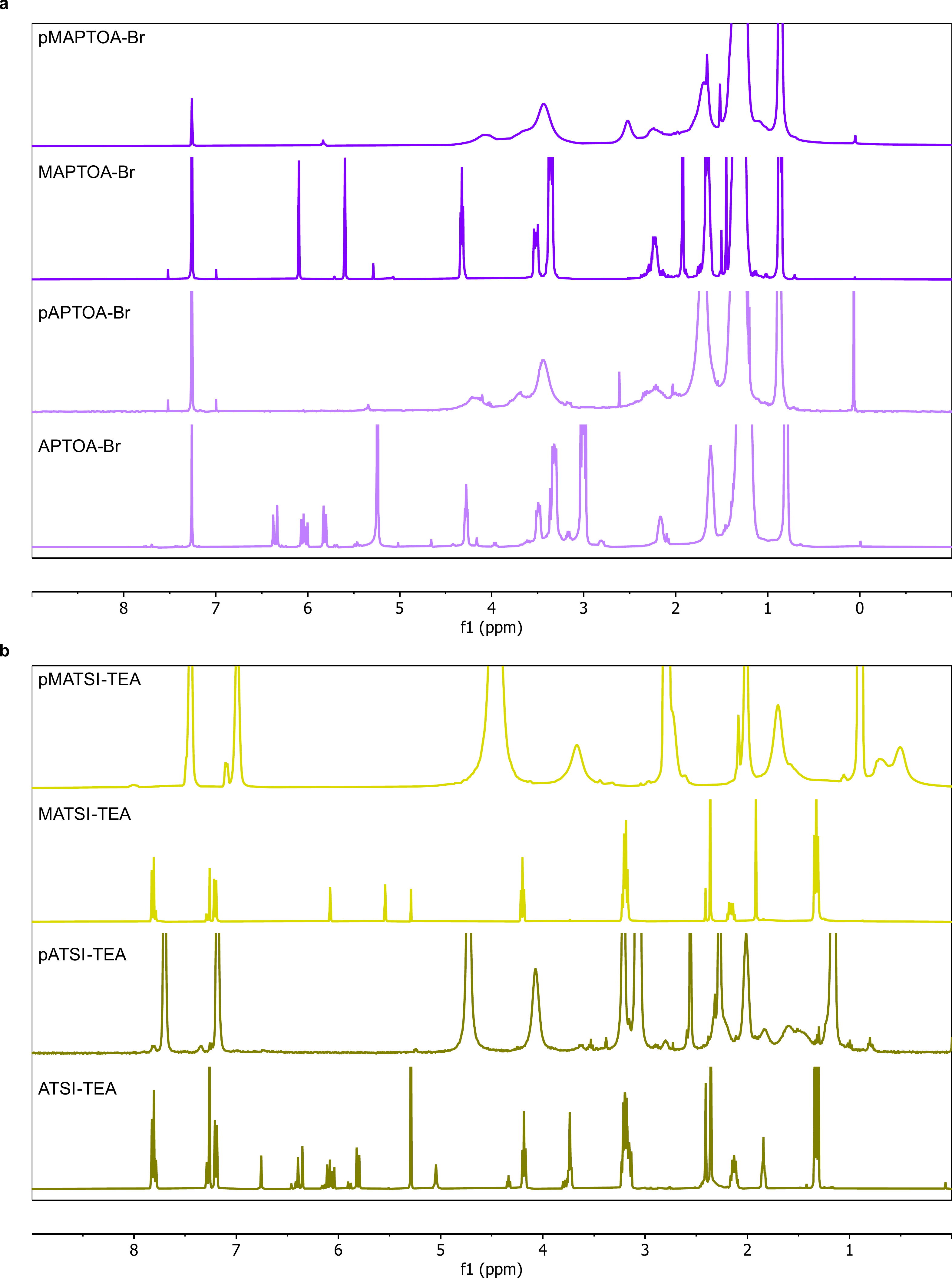}
\caption{\textbf{\textsuperscript{1}H-NMR spectra} of \textbf{(A)} all cationic compounds and \textbf{(B)} all anionic compounds. pATSI-TEA was measured in CD\textsubscript{3}OD and pMATSI-TEA was measured in D\textsubscript{2}O, all other spectra were recorded in CDCl\textsubscript{3}. The polymerization was verified by the broadening of the peaks, and the disappearance of the monomer peaks at 5.8 and 6.1 ppm for the methacrylates and three multiplets at 5.8; 6.1; 6.4 ppm for the acrylates.}
\label{ch4_figA_11}
\end{center}
\end{figure}



\clearpage

\section{Supplementary Tables}

\begin{figure}[h!]
\caption{\textbf{The fragility, width of the relaxation spectrum and steepness of the thermal transition for several materials as plotted in \textbf{Figure 3} (in addition to the data reported by Boehmer et al.\cite{boehmer1993nonexponential}),} including compleximers (this work) and vitrimers\cite{montarnal2011silica, elling2020reprocessable, roig2023disulfide, zhu2024catalyst, hong2024vitrimer, denissen2015vinylogous}, covalent adaptable networks\cite{rusayyis2021repcrocessable, elling2020reprocessable}, thermoplastics\cite{wu2009correlations, RN95, sanchis1995dynamic}, polymer ionic liquids\cite{nakamura2013viscoelastic, yokokoji2021viscoelastic}, and ionomers\cite{chen2013ionomer, weiss1991viscoelastic} from literature.}
\begin{center}
\includegraphics[width=\textwidth]{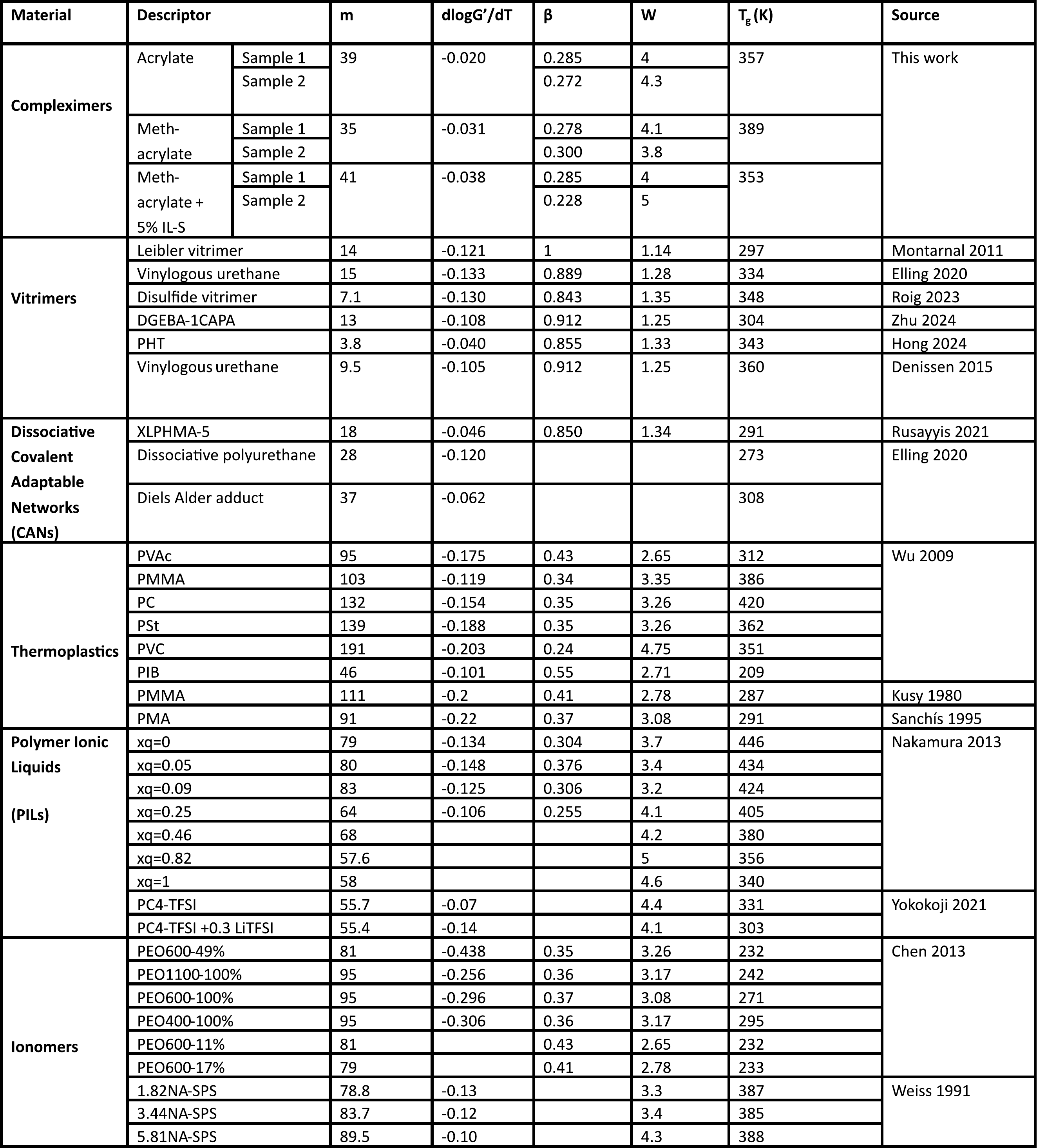} 
\label{ch4_figA__T1}
\end{center}
\end{figure}

\begin{figure}[p!]
\caption{\textbf{The fragility, width of the relaxation spectrum and steepness of the thermal transition for several materials as plotted in \textbf{Figure 3} (in addition to the data reported by Boehmer et al.\cite{boehmer1993nonexponential}),} including ionic liquids\cite{tao2015rheology, shamim2010glass, pogodina2011molecular}, metallic glasses\cite{ikeda2010correlation, qiao2016dynamics}, a network glass\cite{masnik1995complex} and inorganic ionic glasses \cite{matsuda2007calorimetric, kodama2002anharmonicity} from literature.}
\begin{center}
\includegraphics[width=\textwidth]{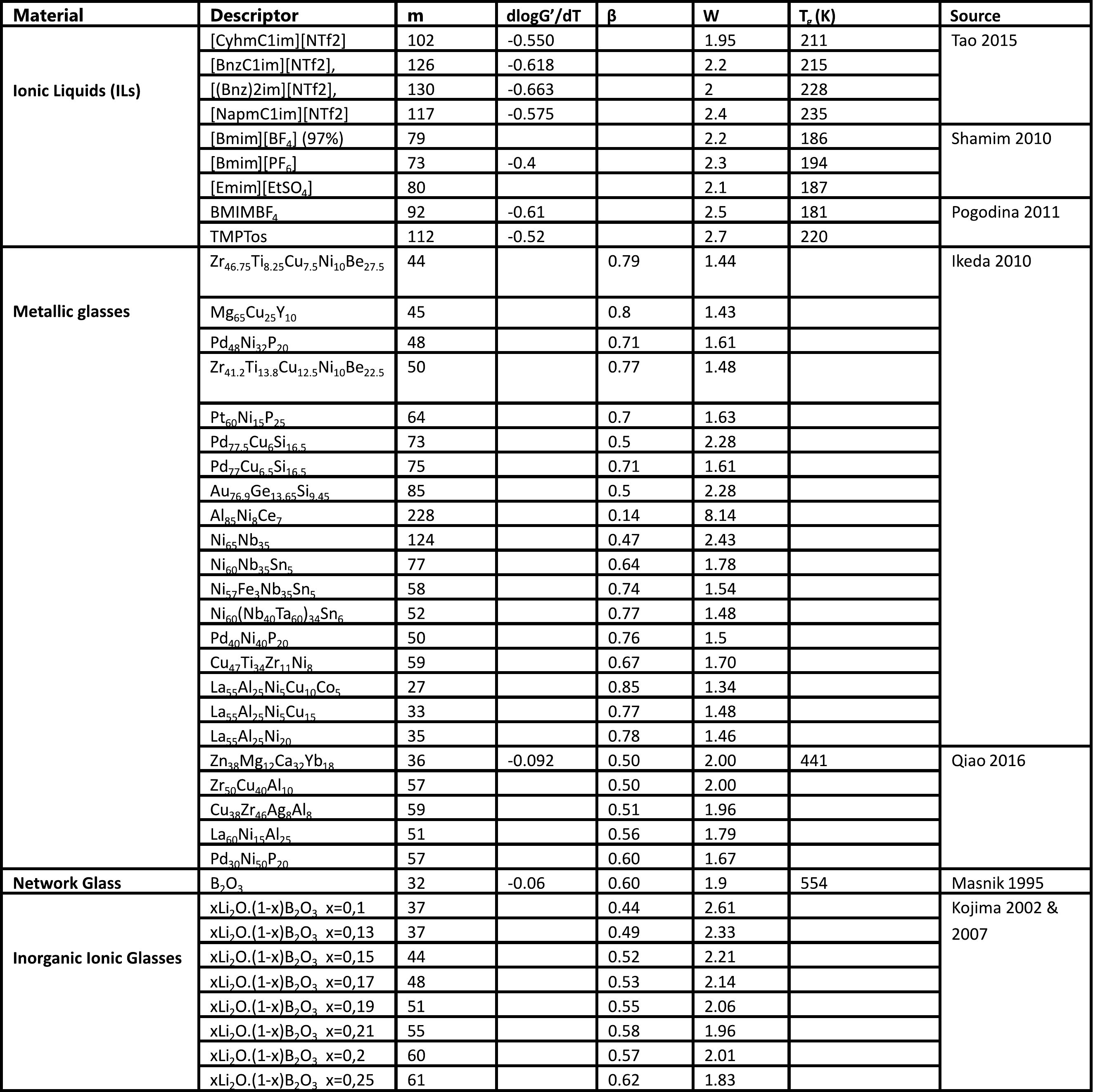} 
\label{ch4_figA__T2}
\end{center}
\end{figure}

\clearpage





\clearpage





\end{appendices}


\begin{thebibliography}{70}
\ifx \bisbn   \undefined \def \bisbn  #1{ISBN #1}\fi
\ifx \binits  \undefined \def \binits#1{#1}\fi
\ifx \bauthor  \undefined \def \bauthor#1{#1}\fi
\ifx \batitle  \undefined \def \batitle#1{#1}\fi
\ifx \bjtitle  \undefined \def \bjtitle#1{#1}\fi
\ifx \bvolume  \undefined \def \bvolume#1{\textbf{#1}}\fi
\ifx \byear  \undefined \def \byear#1{#1}\fi
\ifx \bissue  \undefined \def \bissue#1{#1}\fi
\ifx \bfpage  \undefined \def \bfpage#1{#1}\fi
\ifx \blpage  \undefined \def \blpage #1{#1}\fi
\ifx \burl  \undefined \def \burl#1{\textsf{#1}}\fi
\ifx \doiurl  \undefined \def \doiurl#1{\url{https://doi.org/#1}}\fi
\ifx \betal  \undefined \def \betal{\textit{et al.}}\fi
\ifx \binstitute  \undefined \def \binstitute#1{#1}\fi
\ifx \binstitutionaled  \undefined \def \binstitutionaled#1{#1}\fi
\ifx \bctitle  \undefined \def \bctitle#1{#1}\fi
\ifx \beditor  \undefined \def \beditor#1{#1}\fi
\ifx \bpublisher  \undefined \def \bpublisher#1{#1}\fi
\ifx \bbtitle  \undefined \def \bbtitle#1{#1}\fi
\ifx \bedition  \undefined \def \bedition#1{#1}\fi
\ifx \bseriesno  \undefined \def \bseriesno#1{#1}\fi
\ifx \blocation  \undefined \def \blocation#1{#1}\fi
\ifx \bsertitle  \undefined \def \bsertitle#1{#1}\fi
\ifx \bsnm \undefined \def \bsnm#1{#1}\fi
\ifx \bsuffix \undefined \def \bsuffix#1{#1}\fi
\ifx \bparticle \undefined \def \bparticle#1{#1}\fi
\ifx \barticle \undefined \def \barticle#1{#1}\fi
\bibcommenthead
\ifx \bconfdate \undefined \def \bconfdate #1{#1}\fi
\ifx \botherref \undefined \def \botherref #1{#1}\fi
\ifx \url \undefined \def \url#1{\textsf{#1}}\fi
\ifx \bchapter \undefined \def \bchapter#1{#1}\fi
\ifx \bbook \undefined \def \bbook#1{#1}\fi
\ifx \bcomment \undefined \def \bcomment#1{#1}\fi
\ifx \oauthor \undefined \def \oauthor#1{#1}\fi
\ifx \citeauthoryear \undefined \def \citeauthoryear#1{#1}\fi
\ifx \endbibitem  \undefined \def \endbibitem {}\fi
\ifx \bconflocation  \undefined \def \bconflocation#1{#1}\fi
\ifx \arxivurl  \undefined \def \arxivurl#1{\textsf{#1}}\fi
\csname PreBibitemsHook\endcsname

\bibitem[\protect\citeauthoryear{Angell}{1995}]{angell1995formation}
\begin{barticle}
\bauthor{\bsnm{Angell}, \binits{C.A.}}:
\batitle{Formation of glasses from liquids and biopolymers}.
\bjtitle{Science}
\bvolume{267}(\bissue{5206}),
\bfpage{1924}--\blpage{1935}
(\byear{1995})
\doiurl{10.1126/science.267.5206.1924}
\end{barticle}
\endbibitem

\bibitem[\protect\citeauthoryear{Angell}{1991}]{angell1991relaxation}
\begin{barticle}
\bauthor{\bsnm{Angell}, \binits{C.A.}}:
\batitle{Relaxation in liquids, polymers and plastic crystals — strong/fragile patterns and problems}.
\bjtitle{Journal of Non-Crystalline Solids}
\bvolume{131-133},
\bfpage{13}--\blpage{31}
(\byear{1991})
\doiurl{10.1016/0022-3093(91)90266-9}
\end{barticle}
\endbibitem

\bibitem[\protect\citeauthoryear{Boehmer et~al.}{1993}]{boehmer1993nonexponential}
\begin{barticle}
\bauthor{\bsnm{Boehmer}, \binits{R.}},
\bauthor{\bsnm{Ngai}, \binits{K.L.}},
\bauthor{\bsnm{Angell}, \binits{C.A.}},
\bauthor{\bsnm{Plazek}, \binits{D.J.}}:
\batitle{Nonexponential relaxations in strong and fragile glass formers}.
\bjtitle{Journal of Chemical Physics}
\bvolume{99}(\bissue{5}),
\bfpage{4201}
(\byear{1993})
\end{barticle}
\endbibitem

\bibitem[\protect\citeauthoryear{Plazek and Ngai}{1991}]{plazek1991correlation}
\begin{barticle}
\bauthor{\bsnm{Plazek}, \binits{D.J.}},
\bauthor{\bsnm{Ngai}, \binits{K.L.}}:
\batitle{Correlation of polymer segmental chain dynamics with temperature-dependent time-scale shifts}.
\bjtitle{Macromolecules}
\bvolume{24}(\bissue{5}),
\bfpage{1222}--\blpage{1224}
(\byear{1991})
\end{barticle}
\endbibitem

\bibitem[\protect\citeauthoryear{Hong et~al.}{2011}]{hong2011there}
\begin{barticle}
\bauthor{\bsnm{Hong}, \binits{L.}},
\bauthor{\bsnm{Novikov}, \binits{V.}},
\bauthor{\bsnm{Sokolov}, \binits{A.P.}}:
\batitle{Is there a connection between fragility of glass forming systems and dynamic heterogeneity/cooperativity?}
\bjtitle{Journal of Non-Crystalline Solids}
\bvolume{357}(\bissue{2}),
\bfpage{351}--\blpage{356}
(\byear{2011})
\end{barticle}
\endbibitem

\bibitem[\protect\citeauthoryear{Heuer}{2008}]{heuer2008exploring}
\begin{barticle}
\bauthor{\bsnm{Heuer}, \binits{A.}}:
\batitle{Exploring the potential energy landscape of glass-forming systems: from inherent structures via metabasins to macroscopic transport}.
\bjtitle{Journal of Physics: Condensed Matter}
\bvolume{20}(\bissue{37}),
\bfpage{373101}
(\byear{2008})
\end{barticle}
\endbibitem

\bibitem[\protect\citeauthoryear{Dyre}{2007}]{dyre2007ten}
\begin{barticle}
\bauthor{\bsnm{Dyre}, \binits{J.C.}}:
\batitle{Ten themes of viscous liquid dynamics}.
\bjtitle{Journal of Physics: Condensed Matter}
\bvolume{19}(\bissue{20}),
\bfpage{205105}
(\byear{2007})
\end{barticle}
\endbibitem

\bibitem[\protect\citeauthoryear{Montarnal et~al.}{2011}]{montarnal2011silica}
\begin{barticle}
\bauthor{\bsnm{Montarnal}, \binits{D.}},
\bauthor{\bsnm{Capelot}, \binits{M.}},
\bauthor{\bsnm{Tournilhac}, \binits{F.}},
\bauthor{\bsnm{Leibler}, \binits{L.}}:
\batitle{Silica-like malleable materials from permanent organic networks}.
\bjtitle{Science}
\bvolume{334}(\bissue{6058}),
\bfpage{965}--\blpage{968}
(\byear{2011})
\doiurl{10.1126/science.1212648}
\end{barticle}
\endbibitem

\bibitem[\protect\citeauthoryear{Denissen et~al.}{2016}]{denissen2016vitrimers}
\begin{barticle}
\bauthor{\bsnm{Denissen}, \binits{W.}},
\bauthor{\bsnm{Winne}, \binits{J.M.}},
\bauthor{\bsnm{Du~Prez}, \binits{F.E.}}:
\batitle{Vitrimers: permanent organic networks with glass-like fluidity}.
\bjtitle{Chemical Science}
\bvolume{7}(\bissue{1}),
\bfpage{30}--\blpage{38}
(\byear{2016})
\doiurl{10.1039/C5SC02223A}
\end{barticle}
\endbibitem

\bibitem[\protect\citeauthoryear{Ciarella et~al.}{2019}]{ciarella2019understanding}
\begin{barticle}
\bauthor{\bsnm{Ciarella}, \binits{S.}},
\bauthor{\bsnm{Biezemans}, \binits{R.A.}},
\bauthor{\bsnm{Janssen}, \binits{L.M.C.}}:
\batitle{Understanding, predicting, and tuning the fragility of vitrimeric polymers}.
\bjtitle{Proceedings of the National Academy of Sciences}
\bvolume{116}(\bissue{50}),
\bfpage{25013}--\blpage{25022}
(\byear{2019})
\doiurl{10.1073/pnas.1912571116}
\end{barticle}
\endbibitem

\bibitem[\protect\citeauthoryear{Dalle-Ferrier et~al.}{2016}]{dalle2016many}
\begin{botherref}
\oauthor{\bsnm{Dalle-Ferrier}, \binits{C.}},
\oauthor{\bsnm{Kisliuk}, \binits{A.}},
\oauthor{\bsnm{Hong}, \binits{L.}},
\oauthor{\bsnm{Carini}, \binits{G.}},
\oauthor{\bsnm{D’Angelo}, \binits{G.}},
\oauthor{\bsnm{Alba-Simionesco}, \binits{C.}},
\oauthor{\bsnm{Novikov}, \binits{V.}},
\oauthor{\bsnm{Sokolov}, \binits{A.}}:
Why many polymers are so fragile: A new perspective.
The Journal of chemical physics
\textbf{145}(15)
(2016)
\end{botherref}
\endbibitem

\bibitem[\protect\citeauthoryear{Saika-Voivod et~al.}{2001}]{saika2001fragile}
\begin{barticle}
\bauthor{\bsnm{Saika-Voivod}, \binits{I.}},
\bauthor{\bsnm{Poole}, \binits{P.H.}},
\bauthor{\bsnm{Sciortino}, \binits{F.}}:
\batitle{Fragile-to-strong transition and polyamorphism in the energy landscape of liquid silica}.
\bjtitle{Nature}
\bvolume{412}(\bissue{6846}),
\bfpage{514}--\blpage{517}
(\byear{2001})
\end{barticle}
\endbibitem

\bibitem[\protect\citeauthoryear{Debenedetti and Stillinger}{2001}]{debenedetti2001supercooled}
\begin{barticle}
\bauthor{\bsnm{Debenedetti}, \binits{P.G.}},
\bauthor{\bsnm{Stillinger}, \binits{F.H.}}:
\batitle{Supercooled liquids and the glass transition}.
\bjtitle{Nature}
\bvolume{410}(\bissue{6825}),
\bfpage{259}--\blpage{267}
(\byear{2001})
\end{barticle}
\endbibitem

\bibitem[\protect\citeauthoryear{Sastry et~al.}{1998}]{sastry1998signatures}
\begin{barticle}
\bauthor{\bsnm{Sastry}, \binits{S.}},
\bauthor{\bsnm{Debenedetti}, \binits{P.G.}},
\bauthor{\bsnm{Stillinger}, \binits{F.H.}}:
\batitle{Signatures of distinct dynamical regimes in the energy landscape of a glass-forming liquid}.
\bjtitle{Nature}
\bvolume{393}(\bissue{6685}),
\bfpage{554}--\blpage{557}
(\byear{1998})
\end{barticle}
\endbibitem

\bibitem[\protect\citeauthoryear{Hall and Wolynes}{2003}]{hall2003microscopic}
\begin{barticle}
\bauthor{\bsnm{Hall}, \binits{R.W.}},
\bauthor{\bsnm{Wolynes}, \binits{P.G.}}:
\batitle{Microscopic theory of network glasses}.
\bjtitle{Physical review letters}
\bvolume{90}(\bissue{8}),
\bfpage{085505}
(\byear{2003})
\end{barticle}
\endbibitem

\bibitem[\protect\citeauthoryear{Ruocco et~al.}{2004}]{ruocco2004landscapes}
\begin{barticle}
\bauthor{\bsnm{Ruocco}, \binits{G.}},
\bauthor{\bsnm{Sciortino}, \binits{F.}},
\bauthor{\bsnm{Zamponi}, \binits{F.}},
\bauthor{\bsnm{De~Michele}, \binits{C.}},
\bauthor{\bsnm{Scopigno}, \binits{T.}}:
\batitle{Landscapes and fragilities}.
\bjtitle{The Journal of chemical physics}
\bvolume{120}(\bissue{22}),
\bfpage{10666}--\blpage{10680}
(\byear{2004})
\end{barticle}
\endbibitem

\bibitem[\protect\citeauthoryear{Betancourt et~al.}{2018}]{pazmino2018string}
\begin{botherref}
\oauthor{\bsnm{Betancourt}, \binits{P.B.A.}},
\oauthor{\bsnm{Starr}, \binits{F.W.}},
\oauthor{\bsnm{Douglas}, \binits{J.F.}}:
String-like collective motion in the $\alpha$-and $\beta$-relaxation of a coarse-grained polymer melt.
The Journal of Chemical Physics
\textbf{148}(10)
(2018)
\end{botherref}
\endbibitem

\bibitem[\protect\citeauthoryear{Furukawa and Tanaka}{2016}]{furukawa2016significant}
\begin{barticle}
\bauthor{\bsnm{Furukawa}, \binits{A.}},
\bauthor{\bsnm{Tanaka}, \binits{H.}}:
\batitle{Significant difference in the dynamics between strong and fragile glass formers}.
\bjtitle{Physical Review E}
\bvolume{94}(\bissue{5}),
\bfpage{052607}
(\byear{2016})
\end{barticle}
\endbibitem

\bibitem[\protect\citeauthoryear{Betancourt et~al.}{2013}]{betancourt2013fragility}
\begin{barticle}
\bauthor{\bsnm{Betancourt}, \binits{B.A.P.}},
\bauthor{\bsnm{Douglas}, \binits{J.F.}},
\bauthor{\bsnm{Starr}, \binits{F.W.}}:
\batitle{Fragility and cooperative motion in a glass-forming polymer--nanoparticle composite}.
\bjtitle{Soft Matter}
\bvolume{9}(\bissue{1}),
\bfpage{241}--\blpage{254}
(\byear{2013})
\end{barticle}
\endbibitem

\bibitem[\protect\citeauthoryear{Vilgis}{1993}]{vilgis1993strong}
\begin{barticle}
\bauthor{\bsnm{Vilgis}, \binits{T.A.}}:
\batitle{Strong and fragile glasses: A powerful classification and its consequences}.
\bjtitle{Physical review B}
\bvolume{47}(\bissue{5}),
\bfpage{2882}
(\byear{1993})
\end{barticle}
\endbibitem

\bibitem[\protect\citeauthoryear{Krausser et~al.}{2015}]{krausser2015interatomic}
\begin{barticle}
\bauthor{\bsnm{Krausser}, \binits{J.}},
\bauthor{\bsnm{Samwer}, \binits{K.H.}},
\bauthor{\bsnm{Zaccone}, \binits{A.}}:
\batitle{Interatomic repulsion softness directly controls the fragility of supercooled metallic melts}.
\bjtitle{Proceedings of the National Academy of Sciences}
\bvolume{112}(\bissue{45}),
\bfpage{13762}--\blpage{13767}
(\byear{2015})
\end{barticle}
\endbibitem

\bibitem[\protect\citeauthoryear{Kohlrausch}{1854}]{kohlrausch1854theorie}
\begin{barticle}
\bauthor{\bsnm{Kohlrausch}, \binits{R.}}:
\batitle{Theorie des elektrischen r{\"u}ckstandes in der leidener flasche}.
\bjtitle{Annalen der Physik}
\bvolume{167}(\bissue{2}),
\bfpage{179}--\blpage{214}
(\byear{1854})
\end{barticle}
\endbibitem

\bibitem[\protect\citeauthoryear{Williams and Watts}{1970}]{williams1970non}
\begin{barticle}
\bauthor{\bsnm{Williams}, \binits{G.}},
\bauthor{\bsnm{Watts}, \binits{D.C.}}:
\batitle{Non-symmetrical dielectric relaxation behaviour arising from a simple empirical decay function}.
\bjtitle{Transactions of the Faraday society}
\bvolume{66},
\bfpage{80}--\blpage{85}
(\byear{1970})
\end{barticle}
\endbibitem

\bibitem[\protect\citeauthoryear{Zhang et~al.}{2022}]{zhang2022angell}
\begin{barticle}
\bauthor{\bsnm{Zhang}, \binits{D.}},
\bauthor{\bsnm{Sun}, \binits{D.}},
\bauthor{\bsnm{Gong}, \binits{X.}}:
\batitle{Angell plot from the potential energy landscape perspective}.
\bjtitle{Physical Review E}
\bvolume{106}(\bissue{6}),
\bfpage{064129}
(\byear{2022})
\end{barticle}
\endbibitem

\bibitem[\protect\citeauthoryear{Gupta and Mauro}{2008}]{gupta2008two}
\begin{barticle}
\bauthor{\bsnm{Gupta}, \binits{P.K.}},
\bauthor{\bsnm{Mauro}, \binits{J.C.}}:
\batitle{Two factors governing fragility: Stretching exponent and configurational entropy}.
\bjtitle{Physical Review E—Statistical, Nonlinear, and Soft Matter Physics}
\bvolume{78}(\bissue{6}),
\bfpage{062501}
(\byear{2008})
\end{barticle}
\endbibitem

\bibitem[\protect\citeauthoryear{Xia and Wolynes}{2001}]{xia2001microscopic}
\begin{barticle}
\bauthor{\bsnm{Xia}, \binits{X.}},
\bauthor{\bsnm{Wolynes}, \binits{P.G.}}:
\batitle{Microscopic theory of heterogeneity and nonexponential relaxations in supercooled liquids}.
\bjtitle{Physical Review Letters}
\bvolume{86}(\bissue{24}),
\bfpage{5526}
(\byear{2001})
\end{barticle}
\endbibitem

\bibitem[\protect\citeauthoryear{Ikeda and Aniya}{2010}]{ikeda2010correlation}
\begin{barticle}
\bauthor{\bsnm{Ikeda}, \binits{M.}},
\bauthor{\bsnm{Aniya}, \binits{M.}}:
\batitle{Correlation between fragility and cooperativity in bulk metallic glass-forming liquids}.
\bjtitle{Intermetallics}
\bvolume{18}(\bissue{10}),
\bfpage{1796}--\blpage{1799}
(\byear{2010})
\end{barticle}
\endbibitem

\bibitem[\protect\citeauthoryear{Ojovan and Tournier}{2021}]{ojavan2021on}
\begin{botherref}
\oauthor{\bsnm{Ojovan}, \binits{M.I.}},
\oauthor{\bsnm{Tournier}, \binits{R.F.}}:
On structural rearrangements near the glass transition temperature in amorphous silica.
Materials (Basel)
\textbf{14}(18)
(2021)
\doiurl{10.3390/ma14185235}
\end{botherref}
\endbibitem

\bibitem[\protect\citeauthoryear{Yang et~al.}{2019}]{yang2019detecting}
\begin{barticle}
\bauthor{\bsnm{Yang}, \binits{Y.}},
\bauthor{\bsnm{Zhang}, \binits{S.}},
\bauthor{\bsnm{Zhang}, \binits{X.}},
\bauthor{\bsnm{Gao}, \binits{L.}},
\bauthor{\bsnm{Wei}, \binits{Y.}},
\bauthor{\bsnm{Ji}, \binits{Y.}}:
\batitle{Detecting topology freezing transition temperature of vitrimers by aie luminogens}.
\bjtitle{Nature Communications}
\bvolume{10}(\bissue{1}),
\bfpage{3165}
(\byear{2019})
\doiurl{10.1038/s41467-019-11144-6}
\end{barticle}
\endbibitem

\bibitem[\protect\citeauthoryear{Nakamura et~al.}{2013}]{nakamura2013viscoelastic}
\begin{barticle}
\bauthor{\bsnm{Nakamura}, \binits{K.}},
\bauthor{\bsnm{Fukao}, \binits{K.}},
\bauthor{\bsnm{Inoue}, \binits{T.}}:
\batitle{Viscoelastic behavior of polymerized ionic liquids with various charge densities}.
\bjtitle{Nihon Reoroji Gakkaishi}
\bvolume{41}(\bissue{1}),
\bfpage{21}--\blpage{27}
(\byear{2013})
\end{barticle}
\endbibitem

\bibitem[\protect\citeauthoryear{Ueno et~al.}{2012}]{ueno2012protic}
\begin{barticle}
\bauthor{\bsnm{Ueno}, \binits{K.}},
\bauthor{\bsnm{Zhao}, \binits{Z.}},
\bauthor{\bsnm{Watanabe}, \binits{M.}},
\bauthor{\bsnm{Angell}, \binits{C.A.}}:
\batitle{Protic ionic liquids based on decahydroisoquinoline: Lost superfragility and ionicity-fragility correlation}.
\bjtitle{The Journal of Physical Chemistry B}
\bvolume{116}(\bissue{1}),
\bfpage{63}--\blpage{70}
(\byear{2012})
\end{barticle}
\endbibitem

\bibitem[\protect\citeauthoryear{Michaels}{1965}]{michaels1965polyelectrolyte}
\begin{barticle}
\bauthor{\bsnm{Michaels}, \binits{A.S.}}:
\batitle{Polyelectrolyte complexes}.
\bjtitle{Industrial and Engineering Chemistry}
\bvolume{57}(\bissue{10}),
\bfpage{32}--\blpage{40}
(\byear{1965})
\doiurl{10.1021/ie50670a007}
\end{barticle}
\endbibitem

\bibitem[\protect\citeauthoryear{Wang and Schlenoff}{2014}]{wang2014the}
\begin{barticle}
\bauthor{\bsnm{Wang}, \binits{Q.}},
\bauthor{\bsnm{Schlenoff}, \binits{J.B.}}:
\batitle{The polyelectrolyte complex/coacervate continuum}.
\bjtitle{Macromolecules}
\bvolume{47}(\bissue{9}),
\bfpage{3108}--\blpage{3116}
(\byear{2014})
\doiurl{10.1021/ma500500q}
\end{barticle}
\endbibitem

\bibitem[\protect\citeauthoryear{Zhang et~al.}{2016}]{zhang2016effect}
\begin{barticle}
\bauthor{\bsnm{Zhang}, \binits{Y.}},
\bauthor{\bsnm{Li}, \binits{F.}},
\bauthor{\bsnm{Valenzuela}, \binits{L.D.}},
\bauthor{\bsnm{Sammalkorpi}, \binits{M.}},
\bauthor{\bsnm{Lutkenhaus}, \binits{J.L.}}:
\batitle{Effect of water on the thermal transition observed in poly(allylamine hydrochloride)–poly(acrylic acid) complexes}.
\bjtitle{Macromolecules}
\bvolume{49}(\bissue{19}),
\bfpage{7563}--\blpage{7570}
(\byear{2016})
\doiurl{10.1021/acs.macromol.6b00742}
\end{barticle}
\endbibitem

\bibitem[\protect\citeauthoryear{van Lange et~al.}{2024}]{vanlange2024moderated}
\begin{barticle}
\bauthor{\bsnm{Lange}, \binits{S.G.M.}},
\bauthor{\bsnm{Brake}, \binits{D.W.}},
\bauthor{\bsnm{Portale}, \binits{G.}},
\bauthor{\bsnm{Palanisamy}, \binits{A.}},
\bauthor{\bsnm{Sprakel}, \binits{J.}},
\bauthor{\bsnm{Gucht}, \binits{J.}}:
\batitle{Moderated ionic bonding for water-free recyclable polyelectrolyte complex materials}.
\bjtitle{Science Advances}
\bvolume{10}(\bissue{2}),
\bfpage{3606}
(\byear{2024})
\doiurl{10.1126/sciadv.adi3606}
\end{barticle}
\endbibitem

\bibitem[\protect\citeauthoryear{Gebbie et~al.}{2017}]{gebbie2017long}
\begin{barticle}
\bauthor{\bsnm{Gebbie}, \binits{M.A.}},
\bauthor{\bsnm{Smith}, \binits{A.M.}},
\bauthor{\bsnm{Dobbs}, \binits{H.A.}},
\bauthor{\bsnm{Lee}, \binits{A.A.}},
\bauthor{\bsnm{Warr}, \binits{G.G.}},
\bauthor{\bsnm{Banquy}, \binits{X.}},
\bauthor{\bsnm{Valtiner}, \binits{M.}},
\bauthor{\bsnm{Rutland}, \binits{M.W.}},
\bauthor{\bsnm{Israelachvili}, \binits{J.N.}},
\bauthor{\bsnm{Perkin}, \binits{S.}},
\bauthor{\bsnm{Atkin}, \binits{R.}}:
\batitle{Long range electrostatic forces in ionic liquids}.
\bjtitle{Chemical Communications}
\bvolume{53}(\bissue{7}),
\bfpage{1214}--\blpage{1224}
(\byear{2017})
\doiurl{10.1039/C6CC08820A}
\end{barticle}
\endbibitem

\bibitem[\protect\citeauthoryear{Wu et~al.}{2009}]{wu2009correlations}
\begin{barticle}
\bauthor{\bsnm{Wu}, \binits{J.}},
\bauthor{\bsnm{Huang}, \binits{G.}},
\bauthor{\bsnm{Qu}, \binits{L.}},
\bauthor{\bsnm{Zheng}, \binits{J.}}:
\batitle{Correlations between dynamic fragility and dynamic mechanical properties of several amorphous polymers}.
\bjtitle{Journal of Non-Crystalline Solids}
\bvolume{355}(\bissue{34}),
\bfpage{1755}--\blpage{1759}
(\byear{2009})
\doiurl{10.1016/j.jnoncrysol.2009.06.013}
\end{barticle}
\endbibitem

\bibitem[\protect\citeauthoryear{Spruijt et~al.}{2013}]{spruijt2013linear}
\begin{barticle}
\bauthor{\bsnm{Spruijt}, \binits{E.}},
\bauthor{\bsnm{Cohen~Stuart}, \binits{M.A.}},
\bauthor{\bsnm{Gucht}, \binits{J.}}:
\batitle{Linear viscoelasticity of polyelectrolyte complex coacervates}.
\bjtitle{Macromolecules}
\bvolume{46}(\bissue{4}),
\bfpage{1633}--\blpage{1641}
(\byear{2013})
\doiurl{10.1021/ma301730n}
\end{barticle}
\endbibitem

\bibitem[\protect\citeauthoryear{Zaccone et~al.}{2014}]{zaccone2014linking}
\begin{barticle}
\bauthor{\bsnm{Zaccone}, \binits{A.}},
\bauthor{\bsnm{Winter}, \binits{H.}},
\bauthor{\bsnm{Siebenb{\"u}rger}, \binits{M.}},
\bauthor{\bsnm{Ballauff}, \binits{M.}}:
\batitle{Linking self-assembly, rheology, and gel transition in attractive colloids}.
\bjtitle{Journal of Rheology}
\bvolume{58}(\bissue{5}),
\bfpage{1219}--\blpage{1244}
(\byear{2014})
\end{barticle}
\endbibitem

\bibitem[\protect\citeauthoryear{Urbain et~al.}{1982}]{urbain1982viscosity}
\begin{barticle}
\bauthor{\bsnm{Urbain}, \binits{G.}},
\bauthor{\bsnm{Bottinga}, \binits{Y.}},
\bauthor{\bsnm{Richet}, \binits{P.}}:
\batitle{Viscosity of liquid silica, silicates and alumino-silicates}.
\bjtitle{Geochimica et cosmochimica acta}
\bvolume{46}(\bissue{6}),
\bfpage{1061}--\blpage{1072}
(\byear{1982})
\end{barticle}
\endbibitem

\bibitem[\protect\citeauthoryear{Huang and McKenna}{2001}]{huang2001new}
\begin{barticle}
\bauthor{\bsnm{Huang}, \binits{D.}},
\bauthor{\bsnm{McKenna}, \binits{G.B.}}:
\batitle{New insights into the fragility dilemma in liquids}.
\bjtitle{The Journal of chemical physics}
\bvolume{114}(\bissue{13}),
\bfpage{5621}--\blpage{5630}
(\byear{2001})
\end{barticle}
\endbibitem

\bibitem[\protect\citeauthoryear{Ljubi{\'c} et~al.}{2014}]{ljubic2014time}
\begin{barticle}
\bauthor{\bsnm{Ljubi{\'c}}, \binits{D.}},
\bauthor{\bsnm{Stamenovi{\'c}}, \binits{M.}},
\bauthor{\bsnm{Smithson}, \binits{C.}},
\bauthor{\bsnm{Nujki{\'c}}, \binits{M.}},
\bauthor{\bsnm{Medo}, \binits{B.}},
\bauthor{\bsnm{Puti{\'c}}, \binits{S.}}:
\batitle{Time-temperature superposition principle: Application of wlf equation in polymer analysis and composites}.
\bjtitle{Za{\v{s}}tita materijala}
\bvolume{55}(\bissue{4}),
\bfpage{395}--\blpage{400}
(\byear{2014})
\end{barticle}
\endbibitem

\bibitem[\protect\citeauthoryear{Dixon}{1990}]{dixon1990specific}
\begin{barticle}
\bauthor{\bsnm{Dixon}, \binits{P.K.}}:
\batitle{Specific-heat spectroscopy and dielectric susceptibility measurements of salol at the glass transition}.
\bjtitle{Physical Review B}
\bvolume{42}(\bissue{13}),
\bfpage{8179}--\blpage{8186}
(\byear{1990})
\doiurl{10.1103/PhysRevB.42.8179}
\end{barticle}
\endbibitem

\bibitem[\protect\citeauthoryear{Meng et~al.}{2022}]{meng2022rheology}
\begin{barticle}
\bauthor{\bsnm{Meng}, \binits{F.}},
\bauthor{\bsnm{Saed}, \binits{M.O.}},
\bauthor{\bsnm{Terentjev}, \binits{E.M.}}:
\batitle{Rheology of vitrimers}.
\bjtitle{Nature Communications}
\bvolume{13}(\bissue{1}),
\bfpage{5753}
(\byear{2022})
\doiurl{10.1038/s41467-022-33321-w}
\end{barticle}
\endbibitem

\bibitem[\protect\citeauthoryear{Elling and Dichtel}{2020}]{elling2020reprocessable}
\begin{barticle}
\bauthor{\bsnm{Elling}, \binits{B.R.}},
\bauthor{\bsnm{Dichtel}, \binits{W.R.}}:
\batitle{Reprocessable cross-linked polymer networks: Are associative exchange mechanisms desirable?}
\bjtitle{ACS Central Science}
\bvolume{6}(\bissue{9}),
\bfpage{1488}--\blpage{1496}
(\byear{2020})
\doiurl{10.1021/acscentsci.0c00567}
\end{barticle}
\endbibitem

\bibitem[\protect\citeauthoryear{Roig et~al.}{2023}]{roig2023disulfide}
\begin{barticle}
\bauthor{\bsnm{Roig}, \binits{A.}},
\bauthor{\bsnm{Agizza}, \binits{M.}},
\bauthor{\bsnm{Serra}, \binits{A.}},
\bauthor{\bsnm{Flor}, \binits{S.}}:
\batitle{Disulfide vitrimeric materials based on cystamine and diepoxy eugenol as bio-based monomers}.
\bjtitle{European Polymer Journal}
\bvolume{194},
\bfpage{112185}
(\byear{2023})
\doiurl{10.1016/j.eurpolymj.2023.112185}
\end{barticle}
\endbibitem

\bibitem[\protect\citeauthoryear{Zhu et~al.}{2024}]{zhu2024catalyst}
\begin{barticle}
\bauthor{\bsnm{Zhu}, \binits{Y.}},
\bauthor{\bsnm{Li}, \binits{W.}},
\bauthor{\bsnm{He}, \binits{Z.}},
\bauthor{\bsnm{Zhang}, \binits{K.}},
\bauthor{\bsnm{Nie}, \binits{X.}},
\bauthor{\bsnm{Fu}, \binits{R.}},
\bauthor{\bsnm{Chen}, \binits{J.}}:
\batitle{Catalyst-free cardanol-based epoxy vitrimers for self-healing, shape memory, and recyclable materials}.
\bjtitle{Polymers}
\bvolume{16}(\bissue{3}),
\bfpage{307}
(\byear{2024})
\end{barticle}
\endbibitem

\bibitem[\protect\citeauthoryear{Hong and Goh}{2024}]{hong2024vitrimer}
\begin{barticle}
\bauthor{\bsnm{Hong}, \binits{Y.}},
\bauthor{\bsnm{Goh}, \binits{M.}}:
\batitle{Vitrimer nanocomposites for highly thermal conducting materials with sustainability}.
\bjtitle{Polymers (Basel)}
\bvolume{16}(\bissue{3}),
\bfpage{365}
(\byear{2024})
\doiurl{10.3390/polym16030365}
\end{barticle}
\endbibitem

\bibitem[\protect\citeauthoryear{Cywar et~al.}{2023}]{cywar2023elastomeric}
\begin{barticle}
\bauthor{\bsnm{Cywar}, \binits{R.M.}},
\bauthor{\bsnm{Ling}, \binits{C.}},
\bauthor{\bsnm{Clarke}, \binits{R.W.}},
\bauthor{\bsnm{Kim}, \binits{D.H.}},
\bauthor{\bsnm{Kneucker}, \binits{C.M.}},
\bauthor{\bsnm{Salvachúa}, \binits{D.}},
\bauthor{\bsnm{Addison}, \binits{B.}},
\bauthor{\bsnm{Hesse}, \binits{S.A.}},
\bauthor{\bsnm{Takacs}, \binits{C.J.}},
\bauthor{\bsnm{Xu}, \binits{S.}},
\bauthor{\bsnm{Demirtas}, \binits{M.U.}},
\bauthor{\bsnm{Woodworth}, \binits{S.P.}},
\bauthor{\bsnm{Rorrer}, \binits{N.A.}},
\bauthor{\bsnm{Johnson}, \binits{C.W.}},
\bauthor{\bsnm{Tassone}, \binits{C.J.}},
\bauthor{\bsnm{Allen}, \binits{R.D.}},
\bauthor{\bsnm{Chen}, \binits{E.Y.-X.}},
\bauthor{\bsnm{Beckham}, \binits{G.T.}}:
\batitle{Elastomeric vitrimers from designer polyhydroxyalkanoates with recyclability and biodegradability}.
\bjtitle{Science Advances}
\bvolume{9}(\bissue{47}),
\bfpage{1735}
(\byear{2023})
\doiurl{10.1126/sciadv.adi1735}
\end{barticle}
\endbibitem

\bibitem[\protect\citeauthoryear{Denissen et~al.}{2015}]{denissen2015vinylogous}
\begin{barticle}
\bauthor{\bsnm{Denissen}, \binits{W.}},
\bauthor{\bsnm{Rivero}, \binits{G.}},
\bauthor{\bsnm{Nicolaÿ}, \binits{R.}},
\bauthor{\bsnm{Leibler}, \binits{L.}},
\bauthor{\bsnm{Winne}, \binits{J.M.}},
\bauthor{\bsnm{Du~Prez}, \binits{F.E.}}:
\batitle{Vinylogous urethane vitrimers}.
\bjtitle{Advanced Functional Materials}
\bvolume{25}(\bissue{16}),
\bfpage{2451}--\blpage{2457}
(\byear{2015})
\doiurl{10.1002/adfm.201404553}
\end{barticle}
\endbibitem

\bibitem[\protect\citeauthoryear{Bin~Rusayyis and Torkelson}{2021}]{rusayyis2021repcrocessable}
\begin{barticle}
\bauthor{\bsnm{Bin~Rusayyis}, \binits{M.A.}},
\bauthor{\bsnm{Torkelson}, \binits{J.M.}}:
\batitle{Reprocessable covalent adaptable networks with excellent elevated-temperature creep resistance: facilitation by dynamic, dissociative bis(hindered amino) disulfide bonds}.
\bjtitle{Polymer Chemistry}
\bvolume{12}(\bissue{18}),
\bfpage{2760}--\blpage{2771}
(\byear{2021})
\doiurl{10.1039/D1PY00187F}
\end{barticle}
\endbibitem

\bibitem[\protect\citeauthoryear{Qiao et~al.}{2016}]{qiao2016dynamics}
\begin{barticle}
\bauthor{\bsnm{Qiao}, \binits{J.}},
\bauthor{\bsnm{Casalini}, \binits{R.}},
\bauthor{\bsnm{Pelletier}, \binits{J.-M.}},
\bauthor{\bsnm{Yao}, \binits{Y.}}:
\batitle{Dynamics of the strong metallic glass zn38mg12ca32yb18}.
\bjtitle{Journal of Non-Crystalline Solids}
\bvolume{447},
\bfpage{85}--\blpage{90}
(\byear{2016})
\end{barticle}
\endbibitem

\bibitem[\protect\citeauthoryear{Chen et~al.}{2013}]{chen2013ionomer}
\begin{barticle}
\bauthor{\bsnm{Chen}, \binits{Q.}},
\bauthor{\bsnm{Tudryn}, \binits{G.J.}},
\bauthor{\bsnm{Colby}, \binits{R.H.}}:
\batitle{Ionomer dynamics and the sticky rouse model}.
\bjtitle{Journal of Rheology}
\bvolume{57}(\bissue{5}),
\bfpage{1441}--\blpage{1462}
(\byear{2013})
\doiurl{10.1122/1.4818868}
\end{barticle}
\endbibitem

\bibitem[\protect\citeauthoryear{Tao and Simon}{2015}]{tao2015rheology}
\begin{barticle}
\bauthor{\bsnm{Tao}, \binits{R.}},
\bauthor{\bsnm{Simon}, \binits{S.L.}}:
\batitle{Rheology of imidazolium-based ionic liquids with aromatic functionality}.
\bjtitle{The Journal of Physical Chemistry B}
\bvolume{119}(\bissue{35}),
\bfpage{11953}--\blpage{11959}
(\byear{2015})
\doiurl{10.1021/acs.jpcb.5b06163}
\end{barticle}
\endbibitem

\bibitem[\protect\citeauthoryear{Matsuda et~al.}{2007}]{matsuda2007calorimetric}
\begin{barticle}
\bauthor{\bsnm{Matsuda}, \binits{Y.}},
\bauthor{\bsnm{Fukawa}, \binits{Y.}},
\bauthor{\bsnm{Matsui}, \binits{C.}},
\bauthor{\bsnm{Ike}, \binits{Y.}},
\bauthor{\bsnm{Kodama}, \binits{M.}},
\bauthor{\bsnm{Kojima}, \binits{S.}}:
\batitle{Calorimetric study of the glass transition dynamics in lithium borate glasses over a wide composition range by modulated dsc}.
\bjtitle{Fluid phase equilibria}
\bvolume{256}(\bissue{1-2}),
\bfpage{127}--\blpage{131}
(\byear{2007})
\end{barticle}
\endbibitem

\bibitem[\protect\citeauthoryear{Kodama and Kojima}{2002}]{kodama2002anharmonicity}
\begin{barticle}
\bauthor{\bsnm{Kodama}, \binits{M.}},
\bauthor{\bsnm{Kojima}, \binits{S.}}:
\batitle{Anharmonicity and fragility in lithium borate glasses}.
\bjtitle{Journal of thermal analysis and calorimetry}
\bvolume{69}(\bissue{3}),
\bfpage{961}--\blpage{970}
(\byear{2002})
\end{barticle}
\endbibitem

\bibitem[\protect\citeauthoryear{Lunkenheimer et~al.}{2023}]{lunkenheimer2023thermal}
\begin{barticle}
\bauthor{\bsnm{Lunkenheimer}, \binits{P.}},
\bauthor{\bsnm{Loidl}, \binits{A.}},
\bauthor{\bsnm{Riechers}, \binits{B.}},
\bauthor{\bsnm{Zaccone}, \binits{A.}},
\bauthor{\bsnm{Samwer}, \binits{K.}}:
\batitle{Thermal expansion and the glass transition}.
\bjtitle{Nature Physics}
\bvolume{19}(\bissue{5}),
\bfpage{694}--\blpage{699}
(\byear{2023})
\end{barticle}
\endbibitem

\bibitem[\protect\citeauthoryear{MacDonald and Roy}{1955}]{macdonald1955vibrational}
\begin{barticle}
\bauthor{\bsnm{MacDonald}, \binits{D.}},
\bauthor{\bsnm{Roy}, \binits{S.}}:
\batitle{Vibrational anharmonicity and lattice thermal properties. ii}.
\bjtitle{Physical Review}
\bvolume{97}(\bissue{3}),
\bfpage{673}
(\byear{1955})
\end{barticle}
\endbibitem

\bibitem[\protect\citeauthoryear{Dyre}{1998}]{dyre1998source}
\begin{barticle}
\bauthor{\bsnm{Dyre}, \binits{J.C.}}:
\batitle{Source of non-arrhenius average relaxation time in glass-forming liquids}.
\bjtitle{Journal of non-crystalline solids}
\bvolume{235},
\bfpage{142}--\blpage{149}
(\byear{1998})
\end{barticle}
\endbibitem

\bibitem[\protect\citeauthoryear{Hecksher and Dyre}{2015}]{hecksher2015review}
\begin{barticle}
\bauthor{\bsnm{Hecksher}, \binits{T.}},
\bauthor{\bsnm{Dyre}, \binits{J.C.}}:
\batitle{A review of experiments testing the shoving model}.
\bjtitle{Journal of Non-Crystalline Solids}
\bvolume{407},
\bfpage{14}--\blpage{22}
(\byear{2015})
\end{barticle}
\endbibitem

\bibitem[\protect\citeauthoryear{Milkus et~al.}{2018}]{milkus2018interpretation}
\begin{barticle}
\bauthor{\bsnm{Milkus}, \binits{R.}},
\bauthor{\bsnm{Ness}, \binits{C.}},
\bauthor{\bsnm{Palyulin}, \binits{V.V.}},
\bauthor{\bsnm{Weber}, \binits{J.}},
\bauthor{\bsnm{Lapkin}, \binits{A.}},
\bauthor{\bsnm{Zaccone}, \binits{A.}}:
\batitle{Interpretation of the vibrational spectra of glassy polymers using coarse-grained simulations}.
\bjtitle{Macromolecules}
\bvolume{51}(\bissue{4}),
\bfpage{1559}--\blpage{1572}
(\byear{2018})
\end{barticle}
\endbibitem

\bibitem[\protect\citeauthoryear{Zaccone}{2020}]{zaccone2020relaxation}
\begin{barticle}
\bauthor{\bsnm{Zaccone}, \binits{A.}}:
\batitle{Relaxation and vibrational properties in metal alloys and other disordered systems}.
\bjtitle{Journal of Physics: Condensed Matter}
\bvolume{32}(\bissue{20}),
\bfpage{203001}
(\byear{2020})
\end{barticle}
\endbibitem

\bibitem[\protect\citeauthoryear{Cui et~al.}{2017}]{cui2017relation}
\begin{barticle}
\bauthor{\bsnm{Cui}, \binits{B.}},
\bauthor{\bsnm{Milkus}, \binits{R.}},
\bauthor{\bsnm{Zaccone}, \binits{A.}}:
\batitle{The relation between stretched-exponential relaxation and the vibrational density of states in glassy disordered systems}.
\bjtitle{Physics Letters A}
\bvolume{381}(\bissue{5}),
\bfpage{446}--\blpage{451}
(\byear{2017})
\end{barticle}
\endbibitem

\bibitem[\protect\citeauthoryear{Wrede et~al.}{2012}]{wrede2012polyelectrolyte}
\begin{barticle}
\bauthor{\bsnm{Wrede}, \binits{M.}},
\bauthor{\bsnm{Ganza}, \binits{V.}},
\bauthor{\bsnm{Bucher}, \binits{J.}},
\bauthor{\bsnm{Straub}, \binits{B.F.}}:
\batitle{Polyelectrolyte gels comprising a lipophilic, cost-effective aluminate as fluorine-free absorbents for chlorinated hydrocarbons and diesel fuel}.
\bjtitle{ACS Applied Materials \& Interfaces}
\bvolume{4}(\bissue{7}),
\bfpage{3453}--\blpage{3458}
(\byear{2012})
\end{barticle}
\endbibitem

\bibitem[\protect\citeauthoryear{Kammakakam et~al.}{2020}]{kammakakam2020tailored}
\begin{barticle}
\bauthor{\bsnm{Kammakakam}, \binits{I.}},
\bauthor{\bsnm{Bara}, \binits{J.E.}},
\bauthor{\bsnm{Jackson}, \binits{E.M.}},
\bauthor{\bsnm{Lertxundi}, \binits{J.}},
\bauthor{\bsnm{Mecerreyes}, \binits{D.}},
\bauthor{\bsnm{Tom{\'e}}, \binits{L.C.}}:
\batitle{Tailored co2-philic anionic poly (ionic liquid) composite membranes: Synthesis, characterization, and gas transport properties}.
\bjtitle{ACS Sustainable Chemistry and Engineering}
\bvolume{8}(\bissue{15}),
\bfpage{5954}--\blpage{5965}
(\byear{2020})
\end{barticle}
\endbibitem

\bibitem[\protect\citeauthoryear{Ghostine et~al.}{2013}]{RN38}
\begin{barticle}
\bauthor{\bsnm{Ghostine}, \binits{R.A.}},
\bauthor{\bsnm{Shamoun}, \binits{R.F.}},
\bauthor{\bsnm{Schlenoff}, \binits{J.B.}}:
\batitle{Doping and diffusion in an extruded saloplastic polyelectrolyte complex}.
\bjtitle{Macromolecules}
\bvolume{46}(\bissue{10}),
\bfpage{4089}--\blpage{4094}
(\byear{2013})
\doiurl{10.1021/ma4004083}
\end{barticle}
\endbibitem

\bibitem[\protect\citeauthoryear{Yang et~al.}{2021}]{yang2021influence}
\begin{barticle}
\bauthor{\bsnm{Yang}, \binits{Z.}},
\bauthor{\bsnm{Xu}, \binits{X.}},
\bauthor{\bsnm{Xu}, \binits{W.-S.}}:
\batitle{Influence of ionic interaction strength on glass formation of an ion-containing polymer melt}.
\bjtitle{Macromolecules}
\bvolume{54}(\bissue{20}),
\bfpage{9587}--\blpage{9601}
(\byear{2021})
\end{barticle}
\endbibitem

\bibitem[\protect\citeauthoryear{Thompson et~al.}{2022}]{thompson_lammps_2022}
\begin{barticle}
\bauthor{\bsnm{Thompson}, \binits{A.P.}},
\bauthor{\bsnm{Aktulga}, \binits{H.M.}},
\bauthor{\bsnm{Berger}, \binits{R.}},
\bauthor{\bsnm{Bolintineanu}, \binits{D.S.}},
\bauthor{\bsnm{Brown}, \binits{W.M.}},
\bauthor{\bsnm{Crozier}, \binits{P.S.}},
\bauthor{\bsnm{Veld}, \binits{P.J.}},
\bauthor{\bsnm{Kohlmeyer}, \binits{A.}},
\bauthor{\bsnm{Moore}, \binits{S.G.}},
\bauthor{\bsnm{Nguyen}, \binits{T.D.}},
\bauthor{\bsnm{Shan}, \binits{R.}},
\bauthor{\bsnm{Stevens}, \binits{M.J.}},
\bauthor{\bsnm{Tranchida}, \binits{J.}},
\bauthor{\bsnm{Trott}, \binits{C.}},
\bauthor{\bsnm{Plimpton}, \binits{S.J.}}:
\batitle{{LAMMPS} - a flexible simulation tool for particle-based materials modeling at the atomic, meso, and continuum scales}.
\bjtitle{Computer Physics Communications}
\bvolume{271},
\bfpage{108171}
(\byear{2022})
\end{barticle}
\endbibitem

\bibitem[\protect\citeauthoryear{Stukowski}{2010}]{ovito}
\begin{botherref}
\oauthor{\bsnm{Stukowski}, \binits{A.}}:
Visualization and analysis of atomistic simulation data with {OVITO}-the {Open} {Visualization} {Tool}.
Modelling and Simulation in Materials Science and Engineering
\textbf{18}(1)
(2010)
\end{botherref}
\endbibitem

\bibitem[\protect\citeauthoryear{Liu et~al.}{2021}]{liu2021effects}
\begin{barticle}
\bauthor{\bsnm{Liu}, \binits{A.Y.}},
\bauthor{\bsnm{Emamy}, \binits{H.}},
\bauthor{\bsnm{Douglas}, \binits{J.F.}},
\bauthor{\bsnm{Starr}, \binits{F.W.}}:
\batitle{Effects of chain length on the structure and dynamics of semidilute nanoparticle--polymer composites}.
\bjtitle{Macromolecules}
\bvolume{54}(\bissue{7}),
\bfpage{3041}--\blpage{3051}
(\byear{2021})
\end{barticle}
\endbibitem

\end{thebibliography}

\AddToHook{enddocument/afteraux}{%
\immediate\write18{
cp output.aux SI.aux
}%
}
\end{document}